\documentclass[12pt]{amsart}
\usepackage{amsmath,amssymb}
\usepackage{thm-restate}
\usepackage{enumerate}
\usepackage{framed}
\usepackage{geometry}
\usepackage[normalem]{ulem}
\usepackage[colorlinks]{hyperref}
\usepackage[foot]{amsaddr}

\usepackage{listings}
\newcommand{\eq}[1]{\hyperref[eq:#1]{(\ref*{eq:#1})}}
\renewcommand{\sec}[1]{\hyperref[sec:#1]{Section~\ref*{sec:#1}}}
\newcommand{\app}[1]{\hyperref[app:#1]{Appendix~\ref*{app:#1}}}
\newcommand{\theo}[1]{\hyperref[thm:#1]{Theorem~\ref*{thm:#1}}}
\newcommand{\lemm}[1]{\hyperref[lem:#1]{Lemma~\ref*{lem:#1}}}
\newcommand{\defin}[1]{\hyperref[defn:#1]{Definition~\ref*{defn:#1}}}
\newcommand{\corr}[1]{\hyperref[cor:#1]{Corollary~\ref*{cor:#1}}}
\newcommand{\fig}[1]{\hyperref[fig:#1]{Figure~\ref*{fig:#1}}}

\def\ringR{\mathrm{R}}
\begin{document}
\newtheorem{thm}{Theorem}[section]
\newtheorem{lem}[thm]{Lemma}
\newtheorem{dfn}[thm]{Definition}
\newtheorem{cor}[thm]{Corollary}
\def\N{{\mathbb N}}
\def\G{{\mathbb G}}
\def\Q{{\mathbb Q}}
\def\F{{\mathbb F}}
\def\R{{\mathbb R}}
\def\C{{\mathbb C}}
\def\P{{\mathbb P}}
\def\Z{{\mathbb Z}}
\def\v{{\mathbf v}}
\def\x{{\mathbf x}}
\def\O{{\mathcal O}}
\def\M{{\mathcal M}}
\def\kbar{{\bar{k}}}
\def\tr{\mbox{Tr}}
\def\id{\mbox{id}}
\def\im{\mbox{im}}
\def\md{\mbox{mod 2}}
\def\qed{{\tiny $\clubsuit$ \normalsize}}

\renewcommand{\theenumi}{\alph{enumi}}

\DeclareRobustCommand{\dg}[1]{{\color{blue}{#1}}}

\title{Exact synthesis of single-qubit unitaries over Clifford-cyclotomic gate sets}
\author{Simon Forest$^{1,2}$\email{simon.forest@ens.fr}}
\author{David Gosset$^{2,3,4}$\email{dngosset@gmail.com}}
\author{ Vadym Kliuchnikov$^{2,5,6}$\email{vadym@microsoft.com}}
\author{ David McKinnon$^7$\email{dmckinnon@uwaterloo.ca}}
\address{$^1$ D\'{e}partement d'Informatique, \'{E}cole Normale Sup\'{e}rieure, Paris}
\address{$^2$ Institute for Quantum Computing, University of Waterloo}
\address{$^3$ Department of Combinatorics and Optimization, University of Waterloo}
\address{$^4$Walter Burke Institute for Theoretical Physics and Institute for Quantum Information and Matter, California Institute of Technology}
\address{$^5$ Quantum Architectures and Computation Group, Microsoft Research}
\address{$^6$ David R. Cheriton School of Computer Science, University of Waterloo}
\address{$^7$ Department of Pure Mathematics, University of Waterloo}

\indent

\begin{abstract}
We generalize an efficient exact synthesis algorithm for single-qubit unitaries over the Clifford+T gate set which was presented by Kliuchnikov, Maslov and Mosca.  Their algorithm takes as input an exactly synthesizable single-qubit unitary--one which can be expressed without error as a product of Clifford and T gates--and outputs a sequence of gates which implements it. The algorithm is optimal in the sense that the length of the sequence, measured by the number of T gates, is smallest possible.  In this paper, for each positive even integer $n$ we consider the ``Clifford-cyclotomic'' gate set consisting of the Clifford group plus a $z$-rotation by $\frac{\pi}{n}$. We present an efficient exact synthesis algorithm which outputs a decomposition using the minimum number of $\frac{\pi}{n}$ $z$-rotations. For the Clifford+T case $n=4$  the group of exactly synthesizable unitaries was shown to be equal to the group of unitaries with entries over the ring $\mathbb{Z}[e^{i\frac{\pi}{n}},1/2]$. We prove that this characterization holds for a handful of other small values of $n$ but the fraction of positive even integers for which it fails to hold is $100\%$.
\end{abstract}

\maketitle

\vspace{-5pt}
\section{Introduction}

It is often convenient to design quantum algorithms using a gate set which includes all single-qubit unitaries. This is justified by the Solovay-Kitaev Theorem which says that any single-qubit unitary can be approximated with error at most $\epsilon$ using a sequence of $\mathrm{polylog (\frac{1}{\epsilon})}$ gates from any finite universal gate set \cite{DS05}. Moreover, the Theorem directly provides an efficient algorithm to compute such a sequence.

However, it is known that the decomposition of a single-qubit gate using the Solovay-Kitaev algorithm can use more gates than is asymptotically necessary.  With this approach, the length of the sequence of gates used to approximate a given unitary to within error $\epsilon$  is $\mathcal{O}\left(\text{log}^c \left(\frac{1}{\epsilon}\right)\right)$ where $c$ is a constant approximately equal to $4$ \cite{DS05}. In contrast a counting argument provides a lower bound of $\Omega\left(\text{log}\left(\frac{1}{\epsilon}\right)\right)$, and universal gate sets  where the shortest possible decomposition achieves this lower bound are known \cite{HRC02,LPS1,LPS2}. In fact this lower bound is achieved for all universal gate sets with algebraic entries \cite{BG08}. Despite this, for many years  it was an open question to find an efficient algorithm which, for some universal gate set, decomposes unitaries using gate sequences of length $\mathcal{O}\left(\text{log}\left(\frac{1}{\epsilon}\right)\right)$.

Recently a new efficient algorithm was introduced which achieves this for the gate set consisting of the Clifford group $\mathcal{C}$ plus the $T=\mathrm{diag} (1,e^{i\frac{\pi}{4}})$ gate \cite{KMM, Selinger}. In order to describe the algorithm it will be convenient to define the group
\[
\mathcal{G}_4=\langle \mathcal{C}, T\rangle
\]
of exactly synthesizable unitaries--those which can be expressed without error as a product of Clifford and T gates. (Here the subscript $4$ indicates that we adjoin a $\pi/4$ $z$-rotation to the Clifford group.) The algorithm is based on the following three ingredients
\begin{enumerate}
\item \textbf{Efficient and optimal exact synthesis algorithm \cite{KMM}}
An efficient algorithm which takes as input a unitary $U\in \mathcal{G}_4$ and outputs a sequence of Clifford and T gates which implements it. The algorithm is optimal in the sense that the number of $T$ gates in the decomposition is the smallest achievable in any decomposition of $U$.
\\

\item \textbf{Number-theoretic characterization of exactly synthesizable unitaries \cite{KMM}}
The group of exactly synthesizable unitaries is equal to the group of $2\times 2$ unitaries with entries in the ring $\ringR_4=\mathbb{Z}[e^{i\frac{\pi}{4}},1/2]$. In other words, we have $\mathcal{G}_4=U_2(\ringR_4)$ where
\[
U_2 (\ringR_4)=\{U\in U(2): U_{ij}\in \mathbb{Z}[e^{i\frac{\pi}{4}},1/2]\}.
\]
(In reference \cite{KMM} the result is stated in terms of the ring $\mathbb{Z}[i,\frac{1}{\sqrt{2}}]$, which is equal to $\mathbb{Z}[e^{i\frac{\pi}{4}},1/2]$).
\\
\item \textbf{Efficient rounding algorithm \cite{Selinger}}
An efficient\footnote{Strictly speaking the algorithm is not proven to be efficient; its efficiency follows from a number-theoretic conjecture.} algorithm which takes as input any single-qubit unitary $V$ and a desired precision $\epsilon$ and outputs a unitary $\tilde{V}\in U_2(\ringR_4)$ which approximates $V$ within error $\epsilon$.
\end{enumerate}

These ingredients are put together in the following way in order to decompose a given unitary $V$. One first uses the rounding algorithm (c) to obtain $\tilde{V}\in U_2(\ringR_4)$ which approximates $V$ within error $\epsilon$. The number-theoretic characterization (b) says that $U_2(\ringR_4)=\mathcal{G}_4$ and so $\tilde{V}$ is exactly synthesizable. One can therefore use the exact synthesis algorithm (a) to obtain an optimal decomposition of $\tilde{V}$ as a product of Clifford and T gates. The resulting decomposition approximates $V$ within error $\epsilon$.  It is shown in reference \cite{Selinger} that this algorithm outputs decompositions with  length scaling optimally as $\mathcal{O}\left(\text{log}\left(\frac{1}{\epsilon}\right)\right)$. Thus it is an optimal improvement over Solovay-Kitaev for the Clifford+T gate set. Similar machinery has since been developed for a few other gate sets \cite{BGS13,KBS14,BRS14}, but each new one seems to pose unique challenges and there is no general theory.

In this work we consider an infinite family of single-qubit gate sets which includes Clifford+T as a special case. For each even $n$, we define the ``Clifford-cyclotomic'' gate set which consists of the Clifford group plus
\[
U_z(\pi/n)= \left(\begin{array}{cc}
1 & 0\\
0 & e^{\frac{i\pi}{n}}
\end{array}\right).
\]
We refer to the group of unitaries which is generated by these gates as the Clifford-cyclotomic group $\mathcal{G}_n$. Throughout this paper we consider the case where $n$ is even. One could also define the groups $\mathcal{G}_n$ for odd $n$, but it is easy to see that if $n$ is odd then $\mathcal{G}_n=\mathcal{G}_{2n}$. In this sense the odd values of $n$ are redundant.

Unitaries from Clifford-cyclotomic gate sets appear in a variety of contexts in quantum computation, often in the special case where $n$ is a power of $2$ (e.g., in the Clifford hierarchy \cite{GC98}, classifying transversal gates for stabilizer codes \cite{AJ14}, and Shor's algorithm \cite{Shor}). Recently some authors have proposed state distillation protocols which can be used to implement $U_z(\pi/2^k)$ fault-tolerantly \cite{LC13,DP14}.

We are interested in whether or not the three ingredients (a), (b), and (c) described above for Clifford+T (the case $n=4$)  can be generalized to Clifford-cyclotomic gate sets corresponding to other values of $n$. Our first result is a generalization of (a):

\begin{enumerate}[(A)]
\item  For each positive even integer $n$, we describe an efficient exact synthesis algorithm which takes as input a unitary $U\in \mathcal{G}_n$ and outputs a sequence of Cliffords and $U_z(\pi/n)$ gates which implements $U$ up to a global phase. The sequence computed by our algorithm uses the minimum possible number of  $U_z(\pi/n)$ gates.
\end{enumerate}

Our exact synthesis algorithm follows a general strategy used in reference \cite{GKMR} for the Clifford+T case. We first show that every $U\in \mathcal{G}_n$ admits a certain canonical form as a product of generators. We then look at the Bloch sphere representation of $U$, which is an SO(3) matrix.  Using basic facts from algebraic number theory we show how the entries of this matrix contain information about the canonical form of $U$--and we describe how it can be recovered efficiently using this information.

Our approach is closely related to a previous body of work studying discrete rotation groups which we believe may find further applications in quantum circuit synthesis \cite{RS1,RS2,CRS}. The mapping from single-qubit unitaries to rotations of the Bloch sphere sends $\mathcal{G}_n$ to a two-generator discrete subgroup of SO(3) of the kind studied in \cite{RS1}. Whereas references \cite{RS1,RS2} are concerned with characterizing the relations in such groups, our goal here is to obtain an explicit decomposition algorithm. However we use many of the same ideas and techniques as reference \cite{RS1} and our algorithm can be viewed as an application and extension of that work.

Our second result is a partial generalization of (b):
\begin{enumerate}[(B)]
\item  Let $\ringR_n=\mathbb{Z}[e^{i\frac{\pi}{n}},1/2]$ and $U_2 (\ringR_n)$ be the group of all $2\times 2$ unitaries with entries in this ring. We prove that $\mathcal{G}_n=U_2(\ringR_n)$ for $n=2,4,6,8,12$ but that this equality holds for a fraction of positive even integers equal to zero.
\end{enumerate}

Thus, for values of $n$ where $\mathcal{G}_n \neq U_2(\ringR_n)$, the analogue of (b) is false, which is an obstacle to directly generalizing the algorithm described above (in the Clifford+T case) to these gate sets.

To establish (B) we again use tools from algebraic number theory.  To prove that $\mathcal{G}_n=U_2(\ringR_n)$ for $n=2,4,6,8,12$ we analytically reduce the problem to checking whether a certain equation can be satisfied and then we use an exhaustive computer search to confirm this. To prove the second part, we consider the subgroups of $z$-rotations in $\mathcal{G}_n$ and $U_2(\ringR_n)$.  We establish a condition which characterizes when these subgroups are equal, and we show that this condition is violated for almost all positive even integers, in the sense that as $N$ approaches infinity, the number of integers in $\{2,4,\ldots,N\}$ satisfying the condition is $o(N)$ (see Corollary~\ref{cor:badcount}).

\noindent We leave the question of generalizing (c) as a direction for future work.

The remainder of this paper is organized as follows. \sec{prelims} contains definitions and basic properties of the objects studied in this paper. In \sec{canon} we describe a canonical form for unitaries in Clifford-cyclotomic groups. Our optimal exact synthesis algorithm, given in \sec{exact}, is based on this canonical form. In \sec{equality} we consider the question of when $\mathcal{G}_n$ is equal to $U_2(\ringR_n)$. The remaining three Sections (\sec{thmproof}, \sec{smallnproof}, and \sec{phthm}) are devoted to proving our results. \app{gloss} is a glossary which contains definitions and facts from algebraic number theory. The first five Sections of this paper can be understood using only the basic algebraic number theory which is reviewed in this glossary.


\section{Clifford-cyclotomic gate sets}\label{sec:prelims}

In this Section we define the single-qubit Clifford group, Clifford-cyclotomic gate sets, and the Clifford-cyclotomic groups generated by them. We define a notion of optimality for a decomposition of a given unitary over one of these gate sets.  Finally, we describe the image of the Clifford-cyclotomic groups under the standard mapping from single-qubit unitaries to rotations of the Bloch sphere.

\subsection{The single-qubit Clifford group}
The single-qubit Pauli operators are
\[
X=\left(\begin{array}{cc}
0 & 1\\
1 & 0
\end{array}\right)\quad Y=\left(\begin{array}{cc}
0 & -i\\
i & 0
\end{array}\right)\quad Z=\left(\begin{array}{cc}
1 & 0\\
0 & -1
\end{array}\right).
\]

The single-qubit Clifford group $\mathcal{C}$ is a group of $2\times 2$ unitaries which map each Pauli
to another Pauli under conjugation up to a possible minus sign, i.e., for each $C\in \mathcal{C}$ and $P\in\{X,Y,Z\}$ we have
\begin{equation}
CPC^{\dagger}=\pm\tilde{P}
\label{eq:clifford}
\end{equation}
for some $\tilde{P}\in\{X,Y,Z\}$. Moreover, any single-qubit unitary with this
property is an element of $\mathcal{C}$ multiplied by some global
phase $e^{i\phi}$.

There are different choices one can make in defining the Clifford group since the conditions described in the previous paragraph only uniquely specify the quotient group of $\mathcal{C}$  modulo its center (i.e., they specify $\mathcal{C}$ modulo its subgroup of global phases, unitaries proportional to the identity). Here we define the single-qubit Clifford group $\mathcal{C}=\langle H_0,S\rangle$ using generators
\begin{equation}
H_0=\frac{1}{2}\left(\begin{array}{cc}
1+i & 1+i\\
1+i & -1-i
\end{array}\right)\qquad S=\left(\begin{array}{cc}
1 & 0\\
0 & i
\end{array}\right).
\label{eq:H0}
\end{equation}
This convention differs from the usual choice, which uses generators $S$ and the Hadamard matrix $H=\zeta_{8}^{-1} H_0$ (here and throughout the paper we write $\zeta_m=e^{\frac{2\pi i}{m}}$ for an $m$th root of unity); however,  since overall global phases are unimportant in quantum computation, the definition used here is operationally equivalent. We shall later prove that, with our definition, a $2\times2$ unitary is an element of $\mathcal{C}$ if and only if it has matrix elements from the ring $\mathbb{Z}[i,\frac{1}{2}]$.

\subsection{Clifford-cyclotomic gate sets}
A Clifford-cyclotomic gate set is obtained by adjoining a $\pi/n$  $z$-rotation to the Clifford group. For each $n\in\{2,4,6,\ldots\}$ we define the Clifford-cyclotomic group generated by this gate set
\begin{equation}
\mathcal{G}_{n}=\langle \mathcal{C}, U_{z}\left(\pi/n\right)\rangle
\label{eq:genset1}
\end{equation}
where
\begin{equation}
U_{z}(\theta)=\left(\begin{array}{cc}
1 & 0\\
0 & e^{i\theta}
\end{array}\right)=e^{i\theta \left(\frac{1-Z}{2}\right)}=\left(\frac{1+e^{i\theta}}{2}\right)+\left(\frac{1-e^{i\theta}}{2}\right)Z.
\label{eq:ztheta}
\end{equation}

Note that each generator, and therefore every unitary in $\mathcal{G}_n$,  has matrix elements over a subring $\ringR_n\subset \mathbb{C}$ of the complex numbers given by
\[
\ringR_n=\mathbb{Z}\left[\zeta_{2n},1/2\right]
\]
(note that $i$ is a power of $\zeta_{2n}$ since $n$ is even).

Although ultimately our goal is to compute decompositions of a given unitary using the generating set \eq{genset1},  it will sometimes be convenient to use other generating sets for $\mathcal{G}_n$. Since $n$ is even we have $S=U_z(\pi/2)=(U_z(\pi/n))^{n/2}$ and so
\begin{equation}
\mathcal{G}_n=\langle H_0, U_z(\pi/n)\rangle.
\label{eq:twogen}
\end{equation}
While this generating set and \eq{genset1} both appear to single out the $z$-axis, we now define another generating set which does not have this property.  For any $p\in \{x,y,z\}$ let $P\in \{X,Y,Z\}$ be the corresponding Pauli operator and define
\[
U_{\pm p} (\theta)=e^{i\theta \left(\frac{1\mp P}{2}\right)}=\left(\frac{1+e^{i\theta}}{2}\right)\pm \left(\frac{1-e^{i\theta}}{2}\right)P.
\]
Then $U_{\pm p} (\theta)=CU_{z} (\theta)C^{\dagger}$ where $C$ is a Clifford satisfying $CZC^{\dagger}=\pm P$. So $\mathcal{G}_n$ also contains $U_{\pm p} (\pi/n)\in \mathcal{G}_n$ for all $p\in \{x,y,z\}$. A more symmetric generating set for $\mathcal{G}_n$ is then
\begin{equation}
\mathcal{G}_n=\langle \mathcal{C}, U_{\pm x}\left(\pi/n\right), U_{\pm y}\left(\pi/n\right),U_{\pm z}\left(\pi/n\right)\rangle.
\label{eq:symgenerators}
\end{equation}

\subsection{Optimal decomposition}\label{sec:opt}

We say that a decomposition of a unitary over a Clifford-cyclotomic gate set is optimal if it uses the minimum number of $\pi/n$ $z$-rotations.

Consider a product of Clifford gates and $z$-rotations which is equal to some unitary $U$ up to a global phase, i.e., an expression
\begin{equation}
e^{i\phi} U=C_1 U_z (s_1\pi/n)C_2 U_z (s_2\pi/n)\ldots C_l U_z (s_l\pi/n)C_{l+1}.
\label{eq:Uuptophase}
\end{equation}
where $C_i \in \mathcal{C}$ for $i\in \{1,\ldots,l+1\}$. The cost of such an expression is considered to be the number of non-Clifford generators used, i.e., $\sum_{i=1}^{l} s_i$. For any $U\in \mathcal{G}_n$ we know there exists such an expression with $\phi=0$. We define $\mathcal{T}_n(U)$ to be the minimal cost among all such expressions (allowing any value of $\phi$).

\begin{dfn}
For any $U\in \mathcal{G}_n$ define $\mathcal{T}_n (U)$ to be the minimal number of $U_z (\pi/n)$ gates required to implement $U$ (up to a possible global phase), i.e., the minimum value of $\sum_{i=1}^{l} s_i$ achievable in any expression of the form \eq{Uuptophase}. If a sequence of gates of the form \eq{Uuptophase} achieves this minimum then it is said to be an optimal decomposition of $U$.
\end{dfn}

As an example which will be useful to us later on, consider decomposing the gates $U_{\pm p}(a\pi/n)$, where $1\leq a <\frac{n}{2}$.  Note that
\[
\mathcal{T}_n \left(U_{\pm p}(a\pi/n)\right)\leq a
\]
since $U_{\pm p}(a\pi/n)$ is equal to $CU_{z}(a\pi/n)C^\dagger$ for some Clifford $C$. Using equation \eq{p_minusp} and the fact that $U_{p}(\pi/2)$ is Clifford, we see that $\mathcal{T}_n \left(U_{\pm p}(a\pi/n)\right)\leq (n/2-a)$ and therefore
\begin{equation}
\mathcal{T}_n \left(U_{\pm p}(a\pi/n)\right)\leq \mathrm{min}(a, n/2-a) \qquad 1\leq a <\frac{n}{2}
\label{eq:Tcount_gen}
\end{equation}
Unsurprisingly, equality holds in the above equation (this follows from more general results we present later).

\subsection{Bloch sphere representation}
It is well known that, modulo global phases, single-qubit unitaries can be viewed as rotations of the Bloch sphere. In this way the group $\mathcal{G}_n$ is mapped to a subgroup of SO(3).

A unitary is mapped to a rotation matrix in the following way. Any traceless Hermitian $2\times2$ matrix can be written as a linear combination of the Pauli matrices with real coefficients. Using this fact we can see that any $2\times 2$ unitary matrix $V$ can be associated
with a $3\times3$ rotation matrix $\widehat{V}$ defined by
\begin{equation}
VPV^{\dagger}=\sum_{\tilde{P}\in\{X,Y,Z\}}\left(\widehat{V}_{P\tilde{P}}\right)\tilde{P}\qquad P\in\{X,Y,Z\},
\label{eq:blochsphere}
\end{equation}
where the rows and columns of $\widehat{V}$ are indexed by the Pauli matrices. One can easily verify that the map $V\rightarrow\widehat{V}$ is a homomorphism from $\mathrm{U(2)}$ to $\mathrm{SO(3)}$ which identifies unitaries that differ by a global phase.

We write $\mathcal{\widehat{C}}$ and $\widehat{\mathcal{G}}_n$ for the images of $\mathcal{C}$ and $\mathcal{G}_n$ respectively.

The finite group $\mathcal{\widehat{C}}$ consists of all signed permutation matrices with determinant 1. We use a slightly overloaded terminogy whereby we refer to elements of $\widehat{\mathcal{C}}$ as well as elements of $\mathcal{C}$ as Cliffords; however, it will always be clear from the context which case we are in.

 The unitaries $U_x(\pi/n),U_y(\pi/n)$ and $U_z(\pi/n)$ map to rotations about the $x,y$ and $z$-axes which we denote by $R_x,R_y$ and $R_z$ respectively (suppressing the dependence on $n$ for notational convenience). Explicitly, we have
\begin{align*}
R_x &= \left(\begin{array}{ccc}
1 & 0 & 0 \\
0 & \cos\pi/n & \sin\pi/n \\
0 & -\sin\pi/n & \cos\pi/n
\end{array}\right)\hspace*{.2in}
R_y = \left(\begin{array}{ccc}
\cos\pi/n & 0 & -\sin\pi/n \\
0 & 1 & 0 \\
\sin\pi/n & 0 & \cos\pi/n
\end{array}\right) \\
R_z &= \left(\begin{array}{ccc}
\cos\pi/n & \sin\pi/n & 0 \\
-\sin\pi/n & \cos\pi/n & 0 \\
0 & 0 & 1 \\
\end{array}\right)
\end{align*}

In fact, the group $\widehat{\mathcal{G}}_n$ is a discrete two-generator subgroup of SO(3) of the type studied in reference \cite{RS1}: it is generated by two rational-angle rotations about orthogonal axes. This follows from \eq{twogen} which implies that $\widehat{\mathcal{G}}_n$ is generated by $R_z$ and $R_x$ (since $\widehat{H}_0=R_x^{n/2}R_z^{n/2}R_x^{n/2}$).


\section{Canonical form}\label{sec:canon}

In this Section we describe a canonical form for unitaries in $\mathcal{G}_n$. This canonical form expresses a unitary as a product of gates from the generating set \eq{symgenerators}. We will later show that each unitary admits only one such canonical form (it is unique), and we will give an efficient algorithm to compute it.  Other unique canonical forms as well as a complete set of relations for the groups $\mathcal{G}_n$ have previously been established in reference \cite{RS1} using similar techniques.  We believe it should be possible to obtain an exact synthesis algorithm along the lines of what we achieve using one of these other canonical forms. We use the canonical form presented below because it directly generalizes one from \cite{GKMR} which was our starting point for the current work.

We shall use the following simple identities involving the generators \eq{symgenerators}. The $\pi/n$ rotation about the $x,y$ or $z$ axis gives a Clifford matrix when raised to the $n/2$th power:
\begin{equation}
(U_p(\pi/n))^{n/2}=U_p(\pi/2)\in \mathcal{C} \qquad p\in \{x,y,z\}.
\label{eq:relation1}
\end{equation}
Because Cliffords map Paulis to Paulis under conjugation (up to $\pm 1$), we also have ``pseudo-commutation'' relations
\begin{equation}
C U_p(a\pi/n)=U_{\pm p'} (a\pi/n) C \qquad C\in \mathcal{C}, \; p,p'\in \{x,y,z\}.
\label{eq:pseudo}
\end{equation}
This relation allows us to move a Clifford past $U_p(a\pi/n)$ by updating the subscript. Finally, note that
\begin{equation}
U_p(a\pi/n)U_{-p}(a\pi/n)=\zeta^a_{2n},
\label{eq:phaserelation}
\end{equation}
which implies
\begin{equation}
U_{p}(a\pi/n)=\zeta_{2n} ^{-b} U_{p}(\pi/2) U_{-p}(b\pi/n) \qquad b=\frac{n}{2}-a.
\label{eq:p_minusp}
\end{equation}

The following decomposition follows from the above identities, and generalizes one given in \cite{GKMR} for the Clifford+T case.

\begin{lem}
\label{lem:U2decomposition}
Suppose $U\in \mathcal{G}_n$. Then there exists a decomposition
\begin{equation}
U=\left(\prod_{i=1}^{m}U_{p_i}(a_i \pi/n)\right) D\label{eq:U_decomp}
\end{equation}
for some nonnegative integer $m$, rotation axes $p_{1},\ldots, p_{m}\in\{x,y,z\}$ satisfying $p_{i}\neq p_{i+1}$, a matrix $D$ equal to a global phase times a Clifford (i.e., $e^{i\phi}D\in \mathcal{C}$ for some $\phi\in \mathbb{R}$) and integers $1\leq a_i<\frac{n}{2}$.
\end{lem}

\noindent{\it Proof:}
We shall prove a slightly stronger result, namely, that such a decomposition exists satisfying
\begin{equation}
\sum_{i=1}^m \mathrm{min}(a_i, n/2-a_i)=\mathcal{T}_n(U).
\label{eq:condition}
\end{equation}

Let $U\in \mathcal{G}_n$ be given. By definition there exists an optimal decomposition of $U$ of the form \eq{Uuptophase}. Letting $M_j=C_1C_2\ldots C_j$ we can then rewrite \eq{Uuptophase} as
\begin{align}
e^{i\phi}U &=\left(M_1 U_z (s_1\pi/n) M_1^{\dagger}\right)\left(M_2 U_z (s_2\pi/n) M_2^{\dagger}\right)\ldots \left(M_l U_z (s_l\pi/n) M_l^{\dagger}\right) M_{l+1}\\
&= \left(\prod_{i=1}^{l}U_{(-1)^{r_i} f_i}(s_i \pi/n)\right) M_{l+1}\label{eq:r_p}
\end{align}
where $r_i\in \{0,1\}$, $f_i\in \{x,y,z\}$, and, since we started from an optimal decomposition
\begin{equation}
\mathcal{T}_n (U)=\sum_{i=1}^{l} s_i.
\label{eq:optim_assumption}
\end{equation}

Suppose there exists some $i$ in \eq{r_p} where $f_i=f_{i+1}$. If this happens then it must be the case that $r_i = r_{i+1}$. To see this, note that if instead $r_i \neq r_{i+1}$ then the product of the two associated terms
\[
U_{(-1)^{r_i} f_i} (s_i \pi/n)U_{(-1)^{r_{i+1}} f_i} (s_{i+1} \pi/n)
\]
is proportional to $U_{\pm f_i} (|s_i-s_{i+1}|\pi/n)$ for some choice of sign (which can be seen using \eq{phaserelation}). Using this fact and equation \eq{Tcount_gen} we see from \eq{r_p} that $\mathcal{T}_n (U)<\sum_{i=1}^{l} s_i$, which contradicts \eq{optim_assumption}. Having reached a contradiction we conclude that $r_i = r_{i+1}$ if $f_i=f_{i+1}$.

We can combine any consecutive terms  in \eq{r_p} which satisfy $f_i=f_{i+1}$, using the fact that
\begin{equation}
U_p (\theta_1)U_p(\theta_2)=U_p(\theta_1+\theta_2).
\label{eq:addingtheta}
\end{equation}
Combining terms in this way we obtain an expression of the form
\begin{equation}
e^{i\phi}U= \left(\prod_{i=1}^{m}U_{(-1)^{b_i} g_i}(\gamma_i \pi/n)\right)M
\label{eq:decomp_almostdone}
\end{equation}
where $g_i\neq g_{i+1}$, $b_i\in \{0,1\}$, $M\in \mathcal{C}$, and
\begin{equation}
\mathcal{T}_n (U)=\sum_{i=1}^{l} s_i=\sum_{i=1}^{m} {\gamma_i}
\label{eq:sumalpha}
\end{equation}
(the second equality follows because we used equation \eq{addingtheta} to combine terms).

Using \eq{Tcount_gen} we see \eq{sumalpha} implies $\gamma_i\leq \frac{n}{4}$ for all $i$. The decomposition \eq{decomp_almostdone} is almost of the desired form, the only difference arising from the possibility that $b_i\neq 0$.  We now describe a procedure which gets rid of the minus signs and transforms a decomposition of the form \eq{decomp_almostdone} into one of the form \eq{U_decomp}. Each of the integers $a_i$ appearing in the resulting decomposition \eq{decomp_almostdone}  will be equal to either $\gamma_i$ or $\kappa_i=\frac{n}{2}-\gamma_i$. Thus \eq{condition} will be satisfied due to \eq{sumalpha}. The rotation axes $p_i$ in the resulting decomposition may be different from the $g_i$ appearing in \eq{decomp_almostdone}, and likewise the matrix $D$ will in general be different from $M$.

To complete the proof we describe this procedure.  Let $j\in \{1,\ldots,m\}$ be minimal such that $b_j=1$ in \eq{decomp_almostdone}. Equation \eq{p_minusp} shows
\[
U_{- g_j}(\gamma_j \pi/n)=\zeta_{2n} ^{\gamma_j}U_{g_j}(\kappa_j \pi/n)C
\]
where $C\in \mathcal{C}$ is a Clifford diagonal in the Pauli $g_j$-basis. We start by replacing $U_{- g_j}(\gamma_j \pi/n)$ in $\eq{decomp_almostdone}$ with the right hand side of this equation. Then, by repeatedly using the pseudocommutation relation \eq{pseudo}, we move the Clifford $C$ to the far right hand side of the expression, redefining $M\leftarrow \zeta_{2n} ^{\gamma_j}CM$, and updating the rotation axes $g_s$  and the signs $b_s$ for $s>j$. It is not hard to see that the condition $g_i \neq g_{i+1}$ will still hold (for all $i$) after this update (to see this, use the fact that $[CAC^\dagger, CBC^\dagger]=0$ implies $[A, B]=0$).  After rearranging  \eq{decomp_almostdone} in this way we have replaced $-g_j$ with $g_j$ so we ensure $b_i=0$ for all $i\leq j$. We may now repeat this step until, after at most $m$ iterations,  we obtain an expression where $b_i=0$ for all $i\leq m$.
\qed


\section{Optimal exact synthesis}\label{sec:exact}

The purpose of this Section is to describe our exact synthesis algorithm for Clifford-cyclotomic gate sets. Our work can be viewed as an algorithmic version of the techniques presented in reference \cite{RS1}.

We use the fact, established below, that the entries of the Bloch sphere representation $\widehat{U}$ of a unitary $U\in \mathcal{G}_n$ provide information about parameters of the canonical form from \lemm{U2decomposition}. Using this relationship we obtain an algorithm which efficiently recovers the canonical form. As we will see, the canonical form can then be straightforwardly transformed into an optimal decomposition over the gate set \eq{genset1}.

We will need to use the definition of an \textit{algebraic integer} as well as the notion of divisibility of algebraic integers. We reproduce these definitions from the algebraic number theory glossary given in \app{gloss}.

\begin{restatable*}{dfn}{algint}
A complex number $c$ is said to be an {\it algebraic number} iff there is a nonzero polynomial $f(x)$ with coefficients in $\Q$ such that $f(c)=0$.  The number $c$ is said to be an {\it algebraic integer} iff the polynomial $f(x)$ can be chosen to have integer coefficients and leading coefficient one (i.e., $f(x)$ is monic).
\end{restatable*}

\begin{restatable*}{dfn}{divisibility}
Let $b$ and $c$ be algebraic integers.  We say that $b$ is {\it divisible} by $c$ iff $c$ is nonzero and $b/c$ is also an algebraic integer.
\end{restatable*}

To understand how algebraic integers will be useful to us, consider the $a$th power of the matrix $R_z$
\[
R_z^a= \left(\begin{array}{ccc}
\cos\pi a/n & \sin\pi a/n & 0 \\
-\sin\pi a/n & \cos\pi a/n & 0 \\
0 & 0 & 1 \\
\end{array}\right)
\]
We suppose $1\leq a<\frac{n}{2}$ and we now list some properties of this matrix. Our aim is to provide some feeling for what's to come--at this point the reader is not expected to be able to derive these properties (they follow from our more general Theorem given below).

The only nonzero entry of $R_z^a$ which is an algebraic integer is the $1$ in the bottom right. However, there is an algebraic integer $\beta$ (depending only on $n$) such that each of the other seemingly innocuous nonzero entries of this matrix can be rewritten as a quotient
\[
w/\beta^{q_a}
\]
where $w$ is an algebraic integer not divisible by $\beta$, and $q_a$ (``the denominator exponent'') is a nonnegative integer which depends on $a$. Since $1$ is an algebraic integer, we say that its denominator exponent is zero. The pattern of denominator exponents which appears in the matrix $R_z^a$ can therefore be described as follows. There are two rows where the largest denominator exponent appearing in the row is equal to $q_a$. These are the first two rows, those labeled $x$ and $y$. The third row, labeled $z$, has largest denominator exponent equal to $0$.

For example, consider the case $n=12$, with $\beta=2\cos(\pi/4)=\sqrt{2}$.  The matrix $R_z$ can be written as:
\[
R_z= \left(\begin{array}{ccc}
\frac{\sqrt{2-\sqrt{3}}}{2} & \frac{\sqrt{2+\sqrt{3}}}{2} & 0 \\
-\frac{\sqrt{2+\sqrt{3}}}{2} & \frac{\sqrt{2-\sqrt{3}}}{2} & 0 \\
0 & 0 & 1 \\
\end{array}\right)
\]
It turns out that $\sqrt{2\pm\sqrt{3}}$ is an algebraic integer that is not divisible by $\sqrt{2}$, and that $q_a=2$, so that the four entries in the top left of this matrix have the required form of $w/\beta^{q_a}$ for some $w$ not divisible by $\beta$.  The largest denominator exponent appearing the first two rows is $q_a$, as desired, and the denominator exponents in the last row are all zero.

This pattern is the same for $R_z^2$, but we see a slightly different situation for $R_z^3$:
\[
R_z^3= \left(\begin{array}{ccc}
\frac{1}{\sqrt{2}} & \frac{1}{\sqrt{2}} & 0 \\
-\frac{1}{\sqrt{2}} & \frac{1}{\sqrt{2}} & 0 \\
0 & 0 & 1 \\
\end{array}\right)
\]
In this case $q_a=1$, which is reflected by the denominators of $\sqrt{2}$.  But although $q_a$ has changed, the denominator exponents in the first two rows are still $q_a$, and the denominator exponents in the last row are still zero.

Now consider a general element of $\widehat{\mathcal{G}}_n$.  \lemm{U2decomposition} directly implies that the Bloch sphere representation of $U\in \mathcal{G}_n$ can be decomposed as:
\begin{equation}
\widehat{U}=\left(\prod_{i=1}^{m}R_{p_i}^{a_i}\right) \widehat{D}
\label{eq:blochU}
\end{equation}
with $\widehat{D} \in \mathcal{\widehat{C}}$ and parameters $m$, $a_i$, and $p_i$ satisfying the conditions described in the Lemma (they are the same parameters appearing in the decomposition of $U$). We considered the special case $R_z^a$ above and discussed how its denominator exponent pattern is related to $a$. The following Theorem describes how the denominator exponent pattern of $\widehat{U}$ is related to the parameters $m$, $a_i$, and $p_i$.

The statement of the Theorem refers to divisibility and coprimality of algebraic integers.  In this Section we will not need to use the portions of the Theorem having to do with coprimality, so we have not reproduced its definition here.  (We refer the interested reader to \app{gloss}; here we are working in the fraction field of $T_n=\ringR_n\cap \mathbb{R}$ and its ring of integers.)

Let $n=2^k s$ with $k\geq1$ and odd $s$. Define
\begin{equation}
\label{eq:qi_defU}
\beta=\begin{cases}
2 & \text{, if  } k=1\\
2\cos \left(\frac{\pi}{2^k}\right) & \text{, if  } k\geq 2\\
\end{cases}
\qquad
q_a=\begin{cases}
2^{k-1}-2^{k-j} & \text{, if  }\frac{n}{\mathrm{gcd}(a,n)}=2^j\\
2^{k-1} & \text{, otherwise}.
\end{cases}
\end{equation}
Note that $q_a>0$ whenever $1\leq a<\frac{n}{2}$.

\begin{thm}[\textbf{Denominator exponent pattern}]
\label{thm:mainthmU}
Let $U\in \mathcal{G}_n$ and consider its Bloch sphere representation $\widehat{U}$. Suppose $\widehat{U}\notin \widehat{\mathcal{C}}$, and let $\{a_1,\ldots,a_m\}$ and $\{p_1,\ldots,p_m\}$ be the parameters from a decomposition of $U$ of the form given in \lemm{U2decomposition}.  Let $N=\sum_{i}q_{a_i}$. Then
\begin{enumerate}
\item Each nonzero entry of $\widehat{U}$ can be written as a quotient $w/\beta^r$, where $r$ is a nonnegative integer and $w$ is an algebraic integer that is not divisible by $\beta$.
\item The maximum such $r$ which appears in any entry of $\widehat{U}$ is $N$. Exactly two rows of $\widehat{U}$ contain an entry of the form $w/\beta^N$, with $w$ coprime to $\beta$.
\item There is exactly one row of $\widehat{U}$ that does not contain an entry of the form $w/\beta^N$. The maximum value of $r$ appearing in that row is $N-q_{a_1}$; it contains an entry of the form $w/\beta^{N-q_{a_1}}$ where $w$ is coprime to $\beta$. If it is the $i$th row, then $p_1$ is the $i$th entry in the list $\{x,y,z\}$.
\end{enumerate}
\end{thm}

Similar observations about the entries of rotation matrices were used in the proofs given in reference \cite{RS1} (specifically, in the text after Lemma 3); the above Theorem can be viewed as a version of these observations which is suited to our purpose here, which is to develop an algorithm. A proof of \theo{mainthmU} is presented in \sec{thmproof}.

 We now show that the pattern of denominator exponents can be used to infer the decomposition from \lemm{U2decomposition}, which implies that it is unique. We give an algorithmic proof of this fact, that is, our proof directly provides an efficient algorithm to compute the decomposition.

\begin{cor}
\label{cor:uniqueness}
For any $U\in \mathcal{G}_n$ the decomposition from \lemm{U2decomposition} is unique. Moreoever, it satisfies
\begin{equation}
\sum_{i=1}^m \mathrm{min}(a_i, n/2-a_i)=\mathcal{T}_n(U).
\label{eq:suma}
\end{equation}
\end{cor}

\noindent{\it Proof:}
The proof of \lemm{U2decomposition} given in \sec{canon} establishes that there exists a decomposition satisfying \eq{suma}. So to prove the claim we need only show that the decomposition is unique. To this end, it is sufficient to show that the parameters $p_1$ and $a_1$ are uniquely determined by $U$, since, by induction, $p_1,\ldots, p_m$ and $a_1,\ldots, a_m$ are then also uniquely determined, and
\begin{equation}
D=\left(\prod_{i=1}^{m}U_{p_i}(a_i\pi/n)\right)^\dagger U.
\label{eq:C_formula}
\end{equation}

Consider the set of matrices
\[
R^{-b}_q \widehat{U}
\]
where $q\in \{x,y,z\}$ and $1\leq b < \frac{n}{2}$. We have $R^{-b}_q=C R^{\frac{n}{2}-b}_q$ for some Clifford $C\in \widehat{\mathcal{C}}$ (which depends on $q$). Using this fact and equation \eq{blochU} we get
\begin{equation}
R^{-b}_q \widehat{U}=\begin{cases}
C\left(R^{\frac{n}{2}-b}_q \prod_{i=1}^{m}R_{p_i}^{a_i}\right)\widehat{D} & \text{ if $q\neq p_1$} \\
\left(\prod_{i=2}^{m}R_{p_i}^{a_i}\right)\widehat{D}  &  \text{ if $q= p_1$ and $b=a_1$} \\
\left(R^{(a_1-b)}_{p_1}\prod_{i=2}^{m}R_{p_i}^{a_i}\right)\widehat{D}  &  \text{ if $q= p_1$ and $a_1-b>0$} \\
C\left( R^{(a_1+\frac{n}{2}-b)}_{p_1}\prod_{i=2}^{m}R_{p_i}^{a_i}\right)\widehat{D}  &  \text{ if $q= p_1$ and $a_1-b<0$}
\end{cases}
\label{eq:cases_bq}
\end{equation}
Now \theo{mainthmU} implies that each nonzero entry of $R^{-b}_q \widehat{U}$ can be written $\frac{w}{\beta^r}$ where $w$ is not divisible by $\beta$ and $r$ is the denominator exponent.  Let $r_{\max}(b,q)$ be the maximum denominator exponent $r$ which appears in an entry of $R^{-b}_q \widehat{U}$. Now look at equation \eq{cases_bq} and apply the Theorem, keeping in mind that in two of the cases left-multiplication by a Clifford $C$ permutes the entries and multiplies some of them by minus signs, and thus does not alter the set of denominator exponents which appear in the matrix. We see that $r_{\max}(b,q)>r_{\max}(a_1,p_1)$ whenever $(b,q)\neq(a_1,p_1)$. Thus $a_1$ and $p_1$ are uniquely determined by $U$, which completes the proof.
\qed

The above proof directly gives the following algorithm, which takes as input a unitary $U\in \mathcal{G}_n$ and outputs the parameters $m$, $p_1,\ldots,p_m$ and $a_1,\ldots,a_m$ and $D$ which appear in the decomposition from \lemm{U2decomposition}.

\begin{framed}
\noindent \textbf{Algorithm to compute the canonical form}\\
\begin{enumerate}[1]
\item Compute the Bloch sphere representation $\widehat{U}$ and let $M\leftarrow \widehat{U}$ and $i\leftarrow 1$.\\
\item For each $q\in \{x,y,z\}$ and $1\leq b< \frac{n}{2}$, compute the maximum denominator exponent $r_{\rm max}(q,b)$ which appears in an entry of $R_q^{-b}M$. Determine the values $q^\star,b^\star$ for which $r_{\rm max}(q,b)$ is minimal.

\item Set $p_i\leftarrow q^\star $ and $a_i\leftarrow b^\star $ and set $M\leftarrow R_{p_i}^{-{a_i}}M$. If $M$ is a signed permutation matrix then let $m\leftarrow i$ and skip to step 4. Otherwise set $i\leftarrow i+1$ and return to step 2.

\item Compute $D$ using \eq{C_formula}.
\end{enumerate}
\end{framed}

The exact synthesis task we are interested in is to efficiently obtain an optimal decomposition of $U$ (up to a global phase) as a product of Clifford and $U_z(\pi/n)$ gates. The canonical form computed by the above algorithm can be straightforwardly (and efficiently) converted into such a decomposition. We replace each gate $U_{p_i} (a_i\pi/n)$ with a product of Clifford and $U_z(\pi/n)$ gates which is equal to it (up to a global phase). Since the canonical form satisfies \eq{suma}, we are guaranteed that the resulting decomposition of $U$ will be optimal as long as we use $\text{min}(a_i, \frac{n}{2}-a_i)$ non-Clifford gates $U_z(\pi/n)$ to implement $U_{p_i} (a_i\pi/n)$ . We showed how to achieve this in \sec{opt}.

This algorithm requires testing an algebraic integer for divisibility by $\beta$.  We assume that the algebraic integers are given in terms of an integral basis of $\ringR_n$, in which case the divisibility test is straightforward and efficient.  In particular, it can be done in polynomial time in terms of the bit length of the integers.  For example, let $x$ and $y$ be two elements of $\Z[\zeta_{2n}]$.  To see if $x$ is divisible by $y$, first compute the set of Galois conjugates of $y$.  This is easy, since the Galois group of $\Q(\zeta_{2n})$ over $\Q$ acts by permutations on the $2n$th roots of unity.  Multiply all of the nontrivial conjugates of $y$ together to obtain $\gamma\in\Z[\zeta_{2n}]$, and multiply $x$ by $\gamma$.  Clearly $x$ is divisible by $y$ if and only if $x\gamma$ is divisible by $B=y\gamma$ ... but $B\in\Z$, so $x$ is divisible by $y$ if and only if every coefficient of $x\gamma$ in its integral basis representation is divisible by $B$.


\section{For which $n$ is $\mathcal{G}_n$ equal to the group of $2\times2$ unitary matrices with entries in the ring $\ringR_n$?}\label{sec:equality}

Since every element of $\mathcal{G}_n$  has matrix elements in the ring $\ringR_n$, $\mathcal{G}_n$ is a subgroup of
\[
U_2(\ringR_n)=\left\{ V\in U(2) : \; V_{ij}\in \ringR_n\right\},
\]
the group of all $2\times 2 $ unitaries with entries in this ring.  We know of no reason to expect that $U_2(\ringR_n)$ is equal to $\mathcal{G}_n$. Nevertheless, it was shown in reference \cite{KMM} that $\mathcal{G}_4=U_2(\ringR_4)$. This characterization was used in an essential way in the algorithm for approximating single-qubit unitaries over the Clifford+T gate set \cite{Selinger}. Serre has shown that equality holds for $n=4,8$ and does not hold when $n=2^k$ with $k\geq 4$ \cite{Serre}. In this Section we address the question of whether such an equality holds for other values of $n$.  For $n$ where this holds one can hope to extend the approximation strategy used for the Clifford+T gate set.

Our first result is positive--we prove that equality holds for some small values of $n$.
\begin{thm}\label{thm:smalln}
For $n=2,4,6,8,12$ we have $\mathcal{G}_n=U_2(\ringR_n)$.
\end{thm}
The proof of this Theorem is presented in \sec{smallnproof}. The Theorem was previously known to hold in the cases $n=4$ \cite{Serre, KMM}, $n=6$ \cite{Serre,BRS14}, and $n=8$ \cite{Serre} so the result is new only for $n=2$ and $12$.

In the remainder of this Section we show that equality does not hold generically. To prove this we consider the subgroup of $z$-axis rotations in $\mathcal{G}_n$, and the corresponding subgroup in $ U_2 (\ringR_n)$.
\[
D_n =\{U_z(\theta) : U_z(\theta)\in \mathcal{G}_n\}  \qquad \Delta_n =\{U_z(\theta) : U_z(\theta)\in U_2 (\ringR_n)\}.
\]
Of course, if $\mathcal{G}_n$ is equal to $U_2(\ringR_n)$ then these subgroups are equal as well. We characterize exactly those values of $n$ for which $D_n$ equals $\Delta_n$.

We first show that $\mathcal{G}_n$ only contains $z$-rotations by angles which are multiples of $\pi/n$.
\begin{thm}
The subgroup of $z$-axis rotations in $\mathcal{G}_n$ is
\begin{equation}
D_n =\{U_z(\pi j/n) : 0 \leq j \leq 2n-1\}.
\label{eq:Dn}
\end{equation}
\end{thm}
\noindent{\it Proof:}
 Let $U_z(\theta)\in \mathcal{G}_n$ be given, and consider the parameters $m$, $\{p_i\}, \{a_i\}$, and $D$ from the canonical form of this unitary specified in \lemm{U2decomposition}.

First suppose $m=0$. In this case $U_z(\theta)$ is proportional to a Clifford. The only $z$-rotations proportional to Cliffords are
\begin{equation}
\{U_z(0),U_z(\pi),U_z(\pi/2), U_z(3\pi/2)\}.
\label{eq:Cliffordz}
\end{equation}
and all of these unitaries are included in the set \eq{Dn}.

Next suppose $m\geq 1$.  To understand this case we use \theo{mainthmU}. The matrix $U_z(\theta)$ has Bloch sphere representation
\[
\widehat{U_z(\theta)} = \left(\begin{array}{ccc}
\cos \theta & \sin\theta & 0 \\
-\sin \theta & \cos \theta & 0 \\
0 & 0 & 1 \\
\end{array}\right).
\]
In particular, its bottom right entry is equal to $1$. Since $\widehat{U_z(\theta)}$ has a denominator exponent $r=0$ corresponding to its bottom right entry, looking at part (c) of \theo{mainthmU} we see that it must be the case that $N=q_{a_1}$, $m=1$, and $p_1=z$ so that $N-q_{a_1}=0$. In other words
\begin{equation}
U_z(\theta)=U_z(a_1\pi/n)D
\label{eq:simpledecomp}
\end{equation}
for some $a_1\in \{1,\ldots,\frac{n}{2}-1\}$ and $D$ equal to a Clifford up to a global phase. Noting that $D$ is a $z$-rotation (since $D=U_z(\theta-a_1\pi/n)$), we conclude that it is in the set \eq{Cliffordz}. Using this fact and equation \eq{simpledecomp} we see that $U_z(\theta)$ is a power of $U_z(\pi/n)$, which completes the proof.
\qed

Now we turn our attention to $\Delta_n$. From the definition of $U_2 (\ringR_n)$ it is clear that $U_z (\theta)\in U_2(\ringR_n)$ if and only if $e^{i\theta}\in \ringR_n$.  In other words, to characterize the group $\Delta_n$ we have to understand which complex phases $e^{i\theta}$ are in the ring $\ringR_n$ \footnote{The question of identifying the nontrivial phases in the ring $\ringR_n$ is equivalent to determining the rank of the group of $S$-integral points on the unit circle for cyclotomic fields and $S$ equal to the set of places lying over $2$.  The general question of computing the ranks of such curves for arbitrary fields and sets of places was raised in \cite{Le}, at the end of section 2.3.}.

 We prove the following Theorem in \sec{phthm}, using a technique due to Shastri \cite{Sh}.

\begin{restatable}{thm}{phases}
Factor $n=2^ks$, where $s$ is odd, and suppose that there is some positive integer $t$ such that $2^t\equiv -1\pmod s$.  Then the set
\[
S=\{r\in \ringR_n: |r|=1\}
\]
of elements of $\ringR_n$ with complex absolute value $1$ is equal to the set of roots of unity in $\mathbb{Q}(\zeta_{2n})$, that is,
\begin{equation}
S=\{e^{\frac{i\pi j}{n}} : j\in \{0,\ldots,2n-1\}\}
\label{eq:Sprime}
\end{equation}
Conversely, if there is no such positive integer $t$, then the set $S$ contains an element of infinite order.
\label{thm:phasethm}
\end{restatable}

Note that if $S$ contains an element of infinite order then $\Delta_n$ contains a matrix of infinite order and therefore is not equal to $D_n$.

\begin{cor}\label{cor:diag}
Factor $n=2^k s$, where $s$ is odd, and suppose that there is some positive integer $t$ such that $2^t\equiv -1\pmod s$.  Then $\Delta_n=D_n$, and is given by equation \eq{Dn}. Conversely, if there is no such positive integer $t$, then $\Delta_n$ is an infinite group and $G_n \neq U_2(\ringR_n)$.
\end{cor}

To conclude this Section, we now show that almost all even integers do not satisfy the condition in the statement above. This presents an obstacle to generalizing the results of reference \cite{Selinger} to all Clifford-cyclotomic gate sets.
\begin{cor}\label{cor:badcount}
For each positive even integer $N$ let
\[
f_N= \frac{2}{N} \left|\{n\in \{2,4,\ldots, N\} : \mathcal{G}_n=U_2 (\ringR_n)\}\right|
\]
be the fraction of even integers $n$ between $1$ and $N$ for which $ \mathcal{G}_n=U_2 (\ringR_n)$. Then $f_N \to 0$ as $N\to \infty$.
\end{cor}
\noindent{\it Proof:}
Let $g_N$ be the fraction of integers $n\in \{1,2,\ldots \ldots,N\}$  (even and odd)  for which the condition from \corr{diag}  is satisfied, i.e.,
\[
g_N=\frac{1}{N} \left|\{n\in \{1,2,\ldots, N\} : n=2^ks\text{ with $s$ odd, and $\exists$ $t>0$ s.t $2^t\equiv -1\pmod s$} \}\right|.
\]
We shall prove that $g_N\rightarrow 0$ as $N\rightarrow\infty$, which in particular implies that the fraction of \textit{even} integers $n\in \{2,4,\ldots,N\}$ satisfying the condition from \corr{diag} also approaches zero as $N\rightarrow \infty$. Using  \corr{diag} this implies that $f_N\rightarrow 0$ as well.

If $p$ is a prime congruent to $7$ modulo $8$, then $2$ is a perfect square modulo $p$, but $-1$ is not a square modulo $p$, and so $-1$ cannot be a power of $2$ modulo $p$.  If $n$ is a multiple of a prime $p$ congruent to $7$ modulo $8$, then by reducing modulo $p$ we see that $-1$ is also not a power of $2$ modulo $n$.  The proportion of positive integers $n\leq N$ with no prime factor congruent to $7$ modulo $8$ approaches $0$ as $N\to\infty$.  To see this, let $p_i$ denote the $i$th prime number congruent to $7$ modulo $8$, starting from $p_1=7$, and let $r$ be that largest integer such that $p_r\leq N$.  Using the principle of inclusion-exclusion, the number of integers between $1$ and $N$ not divisible by any $p_i$ is exactly:
\[\sum_{I\subseteq\{1,...,r\}}(-1)^{\# I}\left\lfloor\frac{N}{\prod_{i\in I} p_i}\right\rfloor\]
where $\# I$ denotes the number of element of the set $I$.  This sum is at most
\[\sum_{\# I\, \mathrm{even}}\frac{N}{\prod_{i\in I} p_i}-\sum_{\# I\, \mathrm{odd}}\frac{N-1}{\prod_{i\in I} p_i}\]
which is in turn less than or equal to
\[\sum_{I}(-1)^{\# I}\frac{N}{\prod _{i\in I}p_i}+\sum_{I}\frac{1}{\prod_{i\in I} p_i}\]
Each sum can be factored as follows
\[N\prod_{i=1}^r\left(1-\frac{1}{p_i}\right)+\prod_{i=1}^r\left(1+\frac{1}{p_i}\right)\]
By Mertens' Third Theorem (see for example Theorem~429 of \cite{HW}), we have
\[\prod_i\left(1-\frac{1}{p_i}\right)=\Omega\left(\frac{1}{\log N}\right)\]
and
\[\prod_i\left(1+\frac{1}{p_i}\right)=\prod_i\left(1-\frac{1}{p_i^2}\right)/\left(1-\frac{1}{p_i}\right)=O(\log N)\]
since $\prod_i\left(1-\frac{1}{p_i^2}\right)=O(1)$.  Hence
\begin{equation}
g_N\leq  \prod_{i=1}^r\left(1-\frac{1}{p_i}\right)+\frac{1}{N}\prod_{i=1}^r\left(1+\frac{1}{p_i}\right)= \prod_{i=1}^r\left(1-\frac{1}{p_i}\right)+ O\left(\frac{\log N}{N}\right).
\label{eq:gN_bnd}
\end{equation}

To complete the proof we show that the first term on the right-hand side $\rightarrow 0$ as $N\rightarrow \infty$. We have

\[
\log \left(\prod_i\left(1-\frac{1}{p_i}\right) \right)\leq \log\left( \prod_i e^{-1/p_i}\right)=-\sum_i \frac{1}{p_i}.\]

A theorem of Dirichlet (see for example Exercise 7.6 of [3]) implies that this sum is unbounded as $N\to\infty$, and so $\prod_i\left(1-\frac{1}{p_i}\right)\to 0$ as $N\to\infty$. Plugging this into \eq{gN_bnd} we see that $g_N\rightarrow 0$ as well. \qed


\section{Proof of \theo{mainthmU}}\label{sec:thmproof}

Recall that we write $k$ for the number of times that $n$ is divisible by two, i.e. $n=2^ks$ where $s$ is odd. Define $\alpha=2^{2^{1-k}}$, and, for each $1\leq a<\frac{n}{2}$,
\begin{equation}
c_a=\begin{cases}
2\cos(\frac{\pi a}{n}) & \text{, if $\frac{n}{\gcd (a,n)}$ is not a power of $2$} \\
2^{1-2^{1-j}}\cos(\frac{\pi a}{n}) & \text{,  if $\frac{n}{\gcd (a,n)}=2^j$}.
\end{cases}
\end{equation}
Likewise define $s_a$ as above but with $\cos$ replaced by $\sin$. Then
\begin{equation}
\label{eq:cos_sin}
\cos\left(\frac{a \pi}{n}\right)=c_a\alpha^{-q_a} \quad \text{and} \quad \sin\left(\frac{a \pi}{n}\right)=s_a\alpha^{-q_a}.
\end{equation}
where $q_a$ is defined in equation \eq{qi_defU}.

For now, we consider the number field $K=\mathbb{Q}(\{c_a, s_a\; : \: 1\leq a<\frac{n}{2}\}, \alpha)$ and its ring of integers $\mathcal{O}_K$. Later we will see that it is possible to work with a smaller field (the fraction field of $\ringR_n \cap \mathbb{R}$) and its ring of integers.

\begin{lem}
\label{lem:alg_int}
For each $1\leq a<\frac{n}{2}$, both $c_a$ and $s_a$ are algebraic integers coprime to $\alpha$. Moreover, when $\frac{n}{gcd(a,n)}$ is a power of $2$, both $c_a$ and $s_a$ are units.
\end{lem}
{\it Proof:} \/ The minimal polynomial of $\alpha$ over $\mathbb{Q}$ is $x^{2^{k-1}}-2=0$; hence $\alpha$ is an algebraic integer. Furthermore the norm $N(\alpha\mathcal{O}_K)$ in $\mathcal{O}_K$ is a positive power of two. We now show that $c_a$ and $s_a$ are algebraic integers (and hence in $\mathcal{O}_K$) with $N (c_a \mathcal{O}_K)$ and $N (s_a\mathcal{O}_K)$ both odd and hence relatively prime to $N(\alpha\mathcal{O}_K)$. This implies that $\alpha$ and $c_a$ are coprime and that $\alpha$ and $s_a$ are coprime.

It is not hard to see that $\frac{n}{gcd(a,n)}$ is a power of two if and only if $\frac{n}{gcd(n/2-a,n)}$ is. Using this fact we see that $s_a=c_{(\frac{n}{2}-a)}$, and so we need only consider $c_a$ in the rest of the proof.

First suppose $\frac{n}{\gcd (a,n)}$ is not a power of $2$, so $c_a=\zeta^a_{2n}+\zeta_{2n}^{-a}$, where $\zeta_{2n}$ is a primitive $2n$th root of unity. Let $\mu=\zeta^a_{2n}$.  The norm of $(\mu+\mu^{-1})\mathcal{O}_K$ equals the norm of $(\mu^2+1 )\mathcal{O}_K$, since the norm is multiplicative and the norm of $\mu\mathcal{O}_K$ is one.  The norm of $(\mu^2+ 1)\mathcal{O}_K$ is a power of the constant coefficient in the minimal polynomial for $\mu^2+ 1$ over $\Q$.  This minimal polynomial is $\Phi_d(x-1)$ where $d=\frac{n}{\gcd (a,n)}$ and $\Phi_d$ is the $d$th cyclotomic polynomial, whose roots are exactly the primitive $d$th roots of unity.  It is well known that $\Phi_d(-1)$ is odd if $d$ is not a power of 2, but we give a short proof here for the sake of completeness:

\begin{thm}
If $d\geq 3$ is not a power of $2$, then $\Phi_d(\pm 1)$ is odd.  If $d=2^k$ for $k\geq 2$, then $\Phi_d(\pm 1)=2$.
\end{thm}

\noindent
{\it Proof:} \/ If $d=2^k$ for $k\geq 2$, then since $\Phi_{2^k}(x)=x^{2^{k-1}}+1$, the result immediately follows.  Thus, we assume that $d$ is not a power of $2$.

For any integer $n\geq 2$ and prime $p$, every root of $\Phi_{np}(x)$ is also a root of $\Phi_n(x^p)$.  If $p$ divides $n$, then the degree of $\Phi_{np}(x)$ is $p\varphi(n)$ ($\varphi$ is the Euler $\varphi$-function), which is also the degree of $\Phi_n(x^p)$.  Since every cyclotomic polynomial is monic and has no repeated roots, we see that the two polynomials are identical.

If $p$ does not divide $n$, then the degree of $\Phi_{np}(x)$ is $(p-1)\varphi(n)$, but the roots of $\Phi_n(x)$ are roots of $\Phi_n(x^p)$ that are not roots of $\Phi_{np}(x)$.  By comparing degrees again, we see that $\Phi_{np}(x)=\Phi_n(x^p)/\Phi_n(x)$.

To prove the theorem, we proceed by induction on the number of prime factors of $d$ (counting multiplicity).  If $d$ is prime, then it is odd, and since $\Phi_d(x)=x^{d-1}+\ldots+x+1$, we conclude that $\Phi_d(1)=d$ and $\Phi_d(-1)=1$.  In general, write $d=np$ for some odd prime $p$.  If $n>2$ and $p$ divides $n$, then $\Phi_d(\pm 1)=\Phi_n(\pm 1)$ is odd by induction.  If $n>2$ and $p$ does not divide $n$, then $\Phi_d(\pm 1)=\Phi_n(\pm 1)/\Phi_n(\pm 1)=1$, which is odd. If $n=2$, then $\Phi_d(1)=1$ as before, but $\Phi_d(-1)=\Phi_{2p}(-1)=\Phi_p(1)/\Phi_p(-1)=p$, which is odd.  \qed

\vspace{.1in}

Hence $N((\mu+\mu^{-1})\mathcal{O}_K)$ is also odd, since it is a power of an odd number.

Next suppose  $\frac{n}{\gcd (a,n)}=2^j$ for some $j$.  Note that $1\leq a<\frac{n}{2}$ implies $j\geq 2$. To show that $c_a$ is an algebraic integer it is sufficient to show that it is the root of a monic polynomial with algebraic integer coefficients. Let $y=2\cos(a\pi/n)$  and note that $f(y)=0$ where
\[
f(x)=\underbrace{((x^2-2)^2-2)^2...)^2-2}_{\text{iterated $j-1$ times}}
\]
(which can be seen using the double angle formula and the fact that $\frac{n}{\gcd (a,n)}=2^j$). By a straightforward inductive argument we see that this is a monic polynomial of degree $2^{j-1}$ with constant coefficient $2$ and where every coefficient except the leading one is divisible by two. Write
\[
f(x)=x^{2^{j-1}}+2+2\sum_{i=1}^{2^{j-1}-1} b_i x^i
\]
where each $b_i$ is an integer. Substituting $f(y)=f(c_a2^{2^{1-j}})=0$ and dividing through by $2$ gives
\[
c_a^{2^{j-1}}+1+\sum_{i=1}^{2^{j-1}-1} 2^{\frac{i}{2^{j-1}}} b_i c_a^i=0
\]
which shows that $c_a$ is the root of a monic polynomial with algebraic integer coefficients. Hence $c_a$ is an algebraic integer.  We establish that it is coprime to $\alpha$ by showing that its norm is $1$. First note that $2^{\frac{1}{2^{j-1}}}$ has minimal polynomial $x^{2^{j-1}}-2$ and, since $\frac{n}{\gcd (a,n)}=2^j$, $2\cos(a\pi/n)$ has minimal polynomial $\Phi_{2^j}(x-1)$  (over $\mathbb{Q}$).  Letting $D$ be the degree of the extension $K$ we then have (by multiplicativity of the norm)
\[
2^{\frac{D}{2^{j-1}}}=N(2\cos(a\pi/n)\mathcal{O}_K)=N(2^{\frac{1}{2^{j-1}}}\mathcal{O}_K) N(c_a \mathcal{O}_K)=2^{\frac{D}{2^{j-1}}} N(c_a \mathcal{O}_K).
\]
Hence $N(c_a \mathcal{O}_K)=1$ and therefore $c_a$ is a unit and coprime to $\alpha$.
\qed

The following Theorem differs from \theo{mainthmU} in that $\beta$ is replaced by $\alpha$.

\begin{thm}
\label{thm:mainthm}
Let $U\in \mathcal{G}_n$ and consider its Bloch sphere representation $\widehat{U}$.  Suppose $\widehat{U}\notin \widehat{\mathcal{C}}$, and let $\{a_1,\ldots,a_m\}$ and $\{p_1,\ldots,p_m\}$ be the parameters from a decomposition of $U$ of the form given in \lemm{U2decomposition}.  Let $N=\sum_{i}q_{a_i}$. Then
\begin{enumerate}
\item Each nonzero entry of $\widehat{U}$ can be written as a quotient $w/\alpha^r$, where $r$ is a nonnegative integer and $w$ is an algebraic integer that is not divisible by $\alpha$.
\item The maximum such $r$ which appears in any entry of $\widehat{U}$ is $N$. Exactly two rows of $\widehat{U}$ contain an entry of the form $w/\alpha^N$, with $w$ coprime to $\alpha$.
\item There is exactly one row of $\widehat{U}$ that does not contain an entry of the form $w/\alpha^N$. The maximum value of $r$ appearing in that row is $N-q_{a_1}$; it contains an entry of the form $w/\alpha^{N-q_{a_1}}$ where $w$ is coprime to $\alpha$. If it is the $i$th row, then $p_1$ is the $i$th entry in the list $\{x,y,z\}$.
\end{enumerate}
\end{thm}

\noindent
{\it Proof:} \/ Let a decomposition \eq{U_decomp} for $M=\widehat{U}$ be given, with $m\geq1$.

If $m=1$, then $M=R_{p}^aC$ for some rotation axis $p\in\{x,y,z\}$, Clifford $C$ and power $1\leq a<\frac{n}{2}$.  One row of $M$ will consist entirely of zeroes and ones, in exactly the way described in the last item in the theorem, and the other two rows will have entries that are either zero or $\cos a\pi/n$ or $\sin a\pi/n$. Thus, it suffices to prove that $\cos a\pi/n$ and $\sin a\pi/n$ are of the form $w/\alpha^{q_a}$, where $w$ is an algebraic integer coprime to $\alpha$.  This follows directly from equation \eq{cos_sin} and \lemm{alg_int}.

We now suppose $m\geq 2$ and proceed by induction on $m$.  Let
\[
M=(\prod_{i=1}^mR_{p_i}^{a_i})C=R_{p_1}^{a_1}M'
\]
where $M'=(\prod_{i=2}^mR_{p_i}^{a_i})C$. We assume that $p_1=z$ -- the other two cases are symmetrical.  We know that the matrix $R_z$ has the following form:
\[\left(\begin{array}{rrr} w_1/\alpha^{q_{a_1}} & w_2/\alpha^{q_{a_1}} & 0 \\
w_3/\alpha^{q_{a_1}} & w_4/\alpha^{q_{a_1}} & 0 \\
0 & 0 & 1\end{array}\right)\]
where $w_i$ is an algebraic integer coprime to $\alpha$.  If $r_1$, $r_2$, and $r_3$ are the three rows of $M$ and $r'_1$, $r'_2$, and $r'_3$ are the rows of $M'$, we have:
\begin{align*}
r_1 &= (w_1/\alpha^{q_{a_1}})r'_1+(w_2/\alpha^{q_{a_1}})r'_2 \\
r_2 &= (w_3/\alpha^{q_{a_1}})r'_1+(w_4/\alpha^{q_{a_1}})r'_2 \\
r_3 &= r'_3
\end{align*}
Since the entries in each $r'_i$ can be written in the form $w/\alpha^{N-q_{a_1}}$ for some algebraic integer $w$, it follows that the nonzero entries of each $r_i$ can be written in the form $w'/\alpha^N$ for some algebraic integer $w^\prime$.  It is also immediately evident that the entries in $r_3$ can all be written in the form $w/\alpha^{N-q_{a_1}}$ for some algebraic integer $w$, and that there is an entry of the form $w/\alpha^{N-q_{a_1}}$ for some algebraic integer coprime to $\alpha$.

To complete the proof we show that both $r_1$ and $r_2$ contain entries of the form $v/\alpha^N$ where $v$ is coprime to $\alpha$. There is some entry $e$ in one of $r'_1$ or $r'_2$ that is of the form $w/\alpha^{N-q_{a_1}}$ for some algebraic integer coprime to $\alpha$, and the corresponding entry $f$ in the other row ($r'_2$ or $r'_1$) is of the form $w'/\alpha^{N-q_{a_1}-q_{a_2}}$ for some algebraic integer $w'$ (possibly not coprime to $\alpha$, or even zero).  Any linear combination $(s/\alpha^{q_{a_1}})e+(t/\alpha^{q_{a_1}})f$ (where $s$ and $t$ are algebraic integers coprime to $\alpha$) can be written in the form $v/\alpha^{N}$, where $v=sw+\alpha^{q_{a_2}} tw'$ is an algebraic integer coprime to $\alpha$. To see why $v$ and $\alpha$ are coprime note that
\[
v\mathcal{O}_K+\alpha\mathcal{O}_K=sw\mathcal{O}_K+\alpha^{q_{a_2}} tw'\mathcal{O}_K+ \alpha\mathcal{O}_K=sw\mathcal{O}_K+\alpha\mathcal{O}_K.
\]
Now $sw$ and $\alpha$ are coprime since $s$ and $w$ are both coprime to $\alpha$. Hence
\[
v\mathcal{O}_K+\alpha\mathcal{O}_K=sw\mathcal{O}_K+\alpha\mathcal{O}_K=\mathcal{O}_K
\]
and therefore $v$ and $\alpha$ are coprime.
\qed

To complete the proof of \theo{mainthmU} we have to show that $\alpha$ can be replaced by $\beta$ (given by \eq{qi_defU}) in the statement of \theo{mainthm}. Note that when $k=1$ we have $\alpha=\beta$ so take $k\geq 2$. In this case $n=2^k s$ with $1\leq s<\frac{n}{2}$ and
\[
\frac{\beta}{\alpha}=c_s
\]
where $c_s$ is given by \eq{cos_sin} and $\frac{n}{gcd(s,n)}=2^k$. Applying \lemm{alg_int} we see that $c_s$ is a unit. Since $\alpha/\beta$ is a unit, an algebraic integer is divisible by $\alpha$ if and only if it is divisible by $\beta$. Likewise the ideal generated by $\alpha$ is equal to the ideal generated by $\beta$ and so coprimality to $\alpha$ is equivalent to coprimality to $\beta$. Hence $\alpha$ can be replaced by $\beta$ in the statement of the above Theorem. Finally, note that since every element of $\widehat{\mathcal{G}}_n$ has entries in $\ringR_n\cap \mathbb{R}$ and $\beta$ is also in this ring, the algebraic integers $w$ appearing in the Theorem are in the ring of integers of its field of fractions. We can therefore work in this field of fractions and the corresponding ring of integers \footnote{Note that for the rings in this paper, the notions of divisibility and coprimality do not depend on the choice of ambient ring.  In particular, every extension of rings considered in this paper is integral, so that divisibility and coprimality in the smaller ring are equivalent to the corresponding notions in the larger ring.}.


\section{Proof of \theo{smalln}}\label{sec:smallnproof}

\def\p{{\mathfrak{p}}}
\def\vp#1{v_{\mathfrak{\p}}\left(#1\right)}
\def\abss#1{\left|#1\right|^{2}}
\global\long\def\at#1{\left(#1\right)}
\global\long\def\of#1{\left[#1\right]}
\def\set#1{\left\{  #1\right\}  }
\def\K{{\Q(\zeta_{2n})}}
\def\z{{\Z[\zeta_{2n}]}}

We divide the proof into three parts.
\begin{itemize}
\medskip
\item{\textbf{Part 1}: We show that $U\in U_2(\ringR_n)$ satisfies $U\in \mathcal{G}_n$ if and only if the first column of $U$ is a first column of some element of $\mathcal{G}_n$ (\lemm{column}).}
\medskip
\end{itemize}
After proving Part 1, it remains to prove that any normalized vector $(x,y)^T\in \ringR_n$ (i.e., a vector satisfying $|x|^2+|y|^2=1$) appears as the first column of some unitary in $\mathcal{G}_n$. We prove this as follows:
\medskip
\begin{itemize}
\item{\textbf{Part 2}: We define a \textit{complexity measure} function $\mu(x,y)$ which assigns an integer to every vector $(x,y)\in \ringR_n$. We show that any normalized vector with small enough complexity measure (less than or equal to some value $\mu_n$) appears as the first column of some element of $\mathcal{G}_n$ (\lemm{complexity}).  }
\medskip
\item{\textbf{Part 3}: Finally, we show that the complexity measure of a normalized vector $(x,y)$ can be reduced by applying unitaries from $\mathcal{G}_n$. Specifically, there exists a sequence of unitaries $V_1,\ldots, V_m\in \mathcal{G}_n$ such that
\begin{equation}
\left(\begin{array}{c} x_2 \\ y_2\end{array}\right)=V_m V_{m-1}\ldots V_1  \left(\begin{array}{c} x \\ y\end{array}\right)
\label{eq:unitaryseq}
\end{equation}
satisfies $\mu(x_2,y_2)\leq \mu_n$ (this follows directly from \lemm{reduce}).}
\end{itemize}

Note that parts 2 and 3 together imply that $(x,y)^T$ appears as the first column of some unitary in  $\mathcal{G}_n$. Let $(x_2,y_2)$ and $V_1,\ldots,V_m$ be the vector and unitaries from Part 3. Part 2 guarantees that there exists $W\in \mathcal{G}_n$ with first column $(x_2,y_2)^T$. We see from equation \eq{unitaryseq} that
\[
V_1^\dagger \ldots V_{m-1}^\dagger V_m^\dagger W
\]
has first column given by $(x,y)^T$ and is an element of $\mathcal{G}_n$.

While our results in parts 2 and 3 are specific to the cases $n\in \{2,4,6,8,12\}$,  we establish part 1 under the more general assumption that, writing  $n=2^k s$ with $s$ odd, there exists a positive integer $t$ such that $2^t\equiv -1\pmod s$.

\medskip
\begin{center}
\textbf{Part 1}
\end{center}
\medskip

\begin{lem}
Factor $n=2^k s$ where $s$ is odd.  Suppose there is some positive integer $t$ such that $2^t\equiv -1\pmod s$.  Then $U\in U_2(\ringR_n)$ is an element of  $\mathcal{G}_n$ if and only if the first column of $U$ is a first column of some element of $\mathcal{G}_n$.
\label{lem:column}
\end{lem}

\noindent
{\it Proof:}
Suppose $U\in U_2(\ringR_n)$ has the same first column as some unitary $V\in \mathcal{G}_n$. We shall prove that $U$ is an element of $\mathcal{G}_n$. We can always write

\[
U=\left(\begin{array}{cc}
x & -y^{\ast}e^{i\phi}\\
y & x^{\ast}e^{i\phi}
\end{array}\right)
\qquad
V=\left(\begin{array}{cc}
x & -y^{\ast}e^{i\theta}\\
y & x^{\ast}e^{i\theta}
\end{array}\right)
\]
from which we see that $U=VU_z(\phi-\theta)$. Since $V\in \mathcal{G}_n$ we need only show that $U_z(\phi-\theta)\in \mathcal{G}_n$.   $U$ and $V$ have determinants $e^{i\phi}$ and $e^{i\theta}$ respectively, which implies $e^{i\theta},e^{i\phi} \in \ringR_n$. Applying \theo{phasethm} and using the hypothesis that $t$ exists satisfying $2^t\equiv -1\pmod s$ we see that both $e^{i\theta}$ and $e^{i\phi}$ are powers of $\zeta_{2n}$. So $U_z(\phi-\theta)$ is a power of $U_z(\pi/n)$, and therefore an element of $\mathcal{G}_n$.
 \qed

\medskip
\begin{center}
\textbf{Part 2}
\end{center}
\medskip

We use the notion of $\p$-adic valuation in order to define a complexity measure for vectors with entries over $\ringR_n$.
\begin{dfn}
Let $I$ be a fractional ideal in the ring of integers $\mathcal{O}_K$ of a number field $K$ and let $\p$ be a prime ideal in $\mathcal{O}_K$. If $\p^{m}\prod_{i}\p_{i}^{m_{i}}$ is a factorization of $I$,  then the $\p$-adic valuation of $I$, denoted as $\vp I$, is equal to $m$. For any $x$ from $K$ we define $\vp x=\vp{x\mathcal{O_K}}$. We also use the convention $\vp 0=\infty$.
\end{dfn}
It is well known that the $\p$-adic valuation satisfies
\begin{equation}
\vp {xy} = \vp{x} + \vp{y} \label{eq:prop1},
\end{equation}
and non-negative for any algebraic integer $x$ or integral ideal $I$.

We work in the fraction field of $\ringR_n$, which is the cyclotomic field $\Q(\zeta_{2n})$. Its ring of integers is $\z$. For the values $n=2,4,6,8,12$ which we consider here, there is a unique prime ideal $\p$ in $\z$ which contains $2$ (this follows from Theorem 2.13 in \cite{W}). We use the $\p$-adic valuation with respect to this prime ideal; from now on $\p$ always refers to this ideal. In fact, in the cases $n=2,4,6,8,12$, the ring of integers $\z$ is a principal ideal domain and so $\p$ is generated by a single element of $\z$ which we denote by $\xi_n$.

We now use these special facts about $\p$ to derive some additional properties of the $\p$-adic valuation which hold in our case. Firstly, we establish that $x\in \ringR_n$ is an algebraic integer if and only if it has non-negative $\p$-adic valuation. Let us show that if $x\in \ringR_n$ is not an algebraic integer then it has negative $\p$-adic valuation. In this case $x$ can be written $z/2^k$ for $z$ from $\z$ not divisible by $2$ and positive integer $k$. Note that it must be the case that $\vp{z} < \vp{2}$, because otherwise $z$ would be in $\p^{\vp{2}} \z = 2\z$. This implies that $\vp{x}$ is equal to $(\vp{z} - k\vp{2}) < 0$.

Now we show that the above implies that any $x\in \ringR_n$ can be written
\begin{equation}
x=\frac{y}{{\xi_n}^{-\vp{x}}}.
\label{eq:zdenom}
\end{equation}
with $y\in \z$ and $y\notin \p$. In other words the $\p$-adic valuation is equal to minus the ``denominator exponent'' with respect to $\xi_n$.\footnote{ In~\cite{TR14} the denominator exponent with respect to $\xi_4=1+\zeta_8$ was used to establish results on two qubit circuit synthesis over the Clifford+T gate library. Here we use the notion of $\p$-adic valuation as its properties, as well as algorithms for its calculation, are well known from the computational number theory literature.} To see why \eq{zdenom} holds, define $y=x {\xi_n}^{-\vp{x}}$, find that $\vp{y}=0$ and $y \in \ringR_n$, and conclude that $y$ is in $\z$ but not in $\p$.

Now using \eq{zdenom} we derive the following property which holds for all $x,y\in \ringR_n$:
\begin{equation}
\text{If }\vp {x}<\vp{y}\text{ then } \vp{x+y}=\vp{x}. \label{eq:prop2}
\end{equation}
To see this, write $x=z/\xi_n^{r}$ and $y=w/\xi_n^s$ with $r>s$ and $z,w\notin \p$. Then $x+y=(z+w\xi_n^{r-s})/\xi_n^r$ and $z+w\xi_n^{r-s}\notin \p$.
Finally, note that since $\p$ is the only prime ideal containing $2$ and complex conjugation is an automorphism of $\z$ we have $\p=\p^*$. The fact that $\p$ is a prime ideal containing $2$ implies that $\p^*$ is also a prime ideal containing $2$. They must coincide since $\p$ is the only prime ideal containing $2$. This implies:
\begin{equation}
\vp{\abss{x}}=\vp{x^*} + \vp{x} = 2\vp{x}
\label{eq:prop3}
\end{equation}
for all $x\in \ringR_n$, which follows because $x\z=\p^{m}\prod_{i}\p_{i}^{m_{i}}$ implies $x^{\ast}\z=\p^{m}\prod_{i}(\p^{\ast}_{i})^{m_{i}}$ and no $\p^{\ast}_{i}$ is equal to $\p$.

In the remainder of the proof we will often use the properties of the $\p$-adic valuation described in equations \eq{prop1}, \eq{prop2}, and \eq{prop3}. We now define the complexity measure for vectors with entries in $\ringR_n$.

\begin{dfn}[\textbf{Complexity measure}] Let $n\in \{2,4,6,8,12\}$.  For $x,y\in \ringR_n$ we define
\[
\mu(x,y) = -\min(\vp x,\vp y).
\]
\end{dfn}

Now we show that every normalized vector with a small value of the complexity measure is a first column of some element of $\mathcal{G}_n$. Define
\[
\mu_n =v_\p (i+1).
\]

\begin{lem}
Let $n\in \{2,4,6,8,12\}$. Suppose $(x,y)\in \ringR_n$, $|x|^2+|y|^2=1$, and $\mu (x,y)\leq \mu_n$. Then there exists $V\in \mathcal{G}_n$ with first column $(x,y)^T$.
\label{lem:complexity}
\end{lem}

Let $(x,y)$ be given satisfying the hypotheses of the Lemma.  Then $x'= x(i+1)$ and $y'= y (i+1)$ are elements of $\Z[\zeta_{2n}]$ (and satisfy $|x'|^2+|y'|^2=2$). To see why they are elements of $\Z[\zeta_{2n}]$, use the fact that $\vp{x'}\ge0,\vp{y'}\ge0$ and the fact that an element of $\ringR_n$ is an algebraic integer if and only if it has non-negative $\p$-adic valuation. To prove the Lemma it is therefore sufficient to show that $(x',y')/(i+1)$ is a first column of some element of $\mathcal{G}_n$ whenever $x',y'\in \Z[\zeta_{2n}]$ satisfy $|x'|^2+|y'|^2=2$. In the following we establish that every such pair is described by one of the following three cases
\begin{itemize}
\item \textbf{Case 1} $(x',y')=(\zeta_{2n}^j,\zeta_{2n}^l)$ for some integers $j,l$.
\item \textbf{Case 2} $(x',y')=((i+1)\zeta_{2n}^j,0)$ for some integer $j$.
\item \textbf{Case 3} $(x',y')=(0,(i+1)\zeta_{2n}^j)$ for some integer $j$.
\end{itemize}
Before we justify this classification, let us pause to show that in each of these three cases $(x,y)^T$ is the first column of a unitary $V\in \mathcal{G}_n$. In cases 2 and 3  we can take $V=\zeta_{2n}^j\cdot \mathbb{I}$, which is an element of $\mathcal{G}_n$ (as can be seen from equation \eq{phaserelation}). In case 1, take
\[
V=\left(\begin{array}{cc}
\frac{1}{2}(1-i)\zeta_{2n}^j & \frac{1}{2}(1-i)\zeta_{2n}^{-l} \\
\frac{1}{2}(1-i)\zeta_{2n}^l  &-\frac{1}{2}(1-i)\zeta_{2n}^{-j}\\
\end{array}\right)
=-i\zeta_{2n}^j U_z((l-j)\pi/n) H_0 U_z(-(l+j)\pi/n).
\]
where $H_0$ is given in equation \eq{H0}. Since each term in the above product is in $\mathcal{G}_n$ we have $V\in \mathcal{G}_n$.

It remains to establish the above claimed classification of pairs $(x',y')$. To this end, we use a map defined on elements of $\Z[\zeta_{2n}]$ as
\[
G\at{x} = \sum_{m=1}^{d} \abss{\sigma_m\at{x}}
\]
where $\{\sigma_1,\ldots,\sigma_d\}$ is the Galois group of $\K$.  Note that $G(x)$ is zero if and only if $x$ is zero. The second property of $G$ which is crucial for our proof follows from the inequality between geometric and arithmetic averages:
\[
\at{\abss{N_\Q \at{x}}}^{1/d}=\at{\prod_{m=1}^{d}\abss{\sigma_m\at{x}}}^{1/d}\le\frac{1}{d}\sum_{m=1}^d\abss{\sigma_{m}\at x}=\frac{1}{d}G(x)
\]
where $N_\Q(x)$ is the norm of $x$ over $\Q$ relative to $\Q(\zeta_{2n})$.  The equation $|x'|^2+|y'|^2=2$ implies $\frac{1}{d}G\at{x'} + \frac{1}{d}G\at{y'}$=2 since $2$ is invariant under every element of the Galois group. From the above equation we see that $\frac{1}{d}G(x)$ is greater than or equal to $1$ for non-zero $x$. If neither $x'$ nor $y'$ is equal to zero then this implies that $\at{\frac{1}{d}G\at{x'},\frac{1}{d}G\at{y'}}=(1,1)$ (since they must sum to two and each is at least one). So the only possible values of $\at{\frac{1}{d}G\at{x'},\frac{1}{d}G\at{y'}}$ are $(1,1)$, $(2,0)$ and $(0,2)$. When $\frac{1}{d}G(x')=1$ the inequality between $G(x')$ and $\abss{N_\Q(x')}$ must in fact be an equality because $\abss{N_\Q(x)}\ge1$ for non-zero $x$. Therefore in this case $|\sigma_m \at{x'}|^2=1$ for all $m=1,\ldots,d$, and it is well known that this implies that $x'$ must be equal to $\zeta_{2n}^j$ for some integer $j$. So in the case where $\at{\frac{1}{d}G\at{x'},\frac{1}{d}G\at{y'}}=(1,1)$, both $x'$ and $y'$ are powers of $\zeta_{2n}$.  In the case when $\at{\frac{1}{d}G\at{x'},\frac{1}{d}G\at{y'}}$ are $(2,0)$ or $(0,2)$ we have $|x'|^2=2$ and $y'=0$  or $|y'|^2=2$ and $x'=0$ respectively. Consider the case where $|x'|^2=2$ and $y'=0$ (the other case is symmetric). Then $x'/(i+1)\in \ringR_n$ has absolute value $1$. Note that each $n\in \{2,4,6,8,12\}$ satisfies the hypothesis of \theo{phasethm} (with $s=1,1,3,1,3$ and $t=1,1,1,1,1$ respectively), and applying this Theorem we see that $x'/(i+1)$ is a power of $\zeta_{2n}$.
\qed

\medskip
\begin{center}
\textbf{Part 3}
\end{center}
\medskip
Our goal in this part of the proof is to show that, if $\mu(x,y)>\mu_n$, then the complexity measure can always be reduced by applying a unitary from $\mathcal{G}_n$ (this is \lemm{reduce}). Our strategy is to show that one can work ``modulo 2'' in a sense that we make precise below. In this way we reduce the problem to a finite number of cases which can then be checked on a computer.

Let us now define what we mean by working $\md$. Any $x'\in \z$ can we written using an integral basis of  $\Z[\zeta_{2n}]$ as $\sum_{k=0}^{d-1} x_k \zeta_{2n}^k$, for $d = \varphi(2n)$ (where $\varphi$ is Euler's phi function) and with each $x_k$ an integer. We define
\[
x'\md = \sum_{k=0}^{d-1} \at{x_k\md} \zeta_{2n}^k.
\]
where for an integer $x\in\Z$, $x\md$ denotes $0$ or $1$ in $\Z$, rather than in $\Z/2\Z$.

The $\md $ function has the following basic properties

\begin{align*}
  (x+y) \md &= \at{(x \md )+(y \md)} \md \\
  (xy) \md &= \at{(x \md ) \cdot (y \md)} \md \\
  \at{ (x \md)^\ast } \md &= x^\ast \md  \\
  \abss{x\md} \md &= \abss{x} \md
\end{align*}
which can be checked by writing $x+y,xy,x^{\ast}$ and $\abss{x}=xx^{\ast}$ in terms of the coordinates of $x,y$ in an integral basis.

In the following we shall also repeatedly use the following relation between $\vp{x'}$ and $\vp{x' \md}$:
\begin{equation}
\min\at{\vp{x'},\vp{2}} = \min\at{ \vp{x'\md},\vp{2}} \label{eq:padic}
\end{equation}
To see that above is true first we note that for any $u,v$ from $\Z[\zeta_{2n}]$ such that $u-v \in 2\Z[\zeta_{2n}]$ and $\vp{u} < \vp{2}$ we have $\vp{u}=\vp{v}$; second we note that $x'-\at{x'\md } \in 2\Z[\zeta_{2n}]$.

The following Lemma, which is proven using a computer program, is the key to our proof of \lemm{reduce}.
\begin{lem}\label{lem:finite}
Let $n\in \{2,4,6,8,12\}$. Suppose $(a,b)\in \z \text{ }\mathrm{mod}\text{ }2$ satisfy $\vp{a}=\vp{b}=0$ and $ \abss{a} + \abss{b}\text{ }\mathrm{mod}\text{ }2= 0 $. Then there exists an integer $k\in \{1,\ldots, 2n\}$ such that
\[
\vp{a+\zeta_{2n}^k b}>\vp{2}/2.
\]
\end{lem}
\noindent{\it Proof:}
The set
\[
s_n=\{x\in  \z \text{ }\mathrm{mod}\text{ }2: \vp{x}=0\}
\]
is finite and so one can in principle directly check the statement of the Lemma by exhaustively considering pairs of elements $a,b$ from $s_n$ using a computer. It is possible to simplify things using the following observation. For each $x \in \z$ define $c(x)=\min\at{ x\zeta_{2n}^{k}, k=1,\ldots,2n}$ where the minimum is taken with respect to some ordering (in practice, we use the lexicographic ordering on vectors of integers, applied to the representation of $x$ in an integral basis). Since $\vp{x}=\vp{x\zeta_{2n}^{k}}$ for all $k$ and $\abss{x\zeta_{2n}^k}=\abss{x}$, one can instead exhaustively consider pairs of elements $a,b$ from the smaller set $s'_n = \set{ c(x) : x \in \z \md, \vp{x} = 0}$. In \fig{script} we provide a MAGMA script which verifies the Lemma using this strategy.
\qed
\noindent

Finally, we prove:
\begin{lem}
Let $n\in \{2,4,6,8,12\}$ and suppose $x,y\in \ringR_n$ satisfy $|x|^2+|y|^2=1$ and $\mu(x,y)>\mu_n$. Then there exists an integer $k$ such that
\[
\mu(c_k,d_k) < \mu(x,y). \label{lem:reduce}
\]
where
\[
\left(\begin{array}{c} c_k \\ d_k\end{array}\right)=H_0U_z(\pi k/n)\left(\begin{array}{c} x \\ y\end{array}\right).
\]
\label{lem:reduce}
\end{lem}

\noindent{\it Proof:}
Recall that (since $\z$ is a principal ideal domain for the values of $n$ we consider) $\p$ is generated by a single element $\xi_n$. Define $x'=\xi_n^{\mu(x,y)}x$ and $y'=\xi_n^{\mu(x,y)}y$. Without loss of generality we can assume that $\vp{x} = -\mu(x,y)$ and thus $\vp{x'}=0$. Then
\begin{equation}
\vp{\abss{x'}+\abss{y'}}=\vp{|\xi_n|^{2\mu(x,y)}}=2\mu{(x,y)}>2\mu_n=\vp{2}>0.
\label{eq:absxy}
\end{equation}
where in the last inequality we used the fact that $\p$ contains 2. Now using \eq{prop2}, \eq{prop3}, \eq{absxy}, and the fact that $\vp{|x'|^2}=0$ we see that $\vp{\abss{y'}} = 0$ and hence $\vp{y'} = 0$. Thus $(x,y)=\xi_{n}^{-\mu(x,y)}(x',y')$ with $\vp{x'}=\vp{y'}=0$.

Using this fact we get
\[
\left(\begin{array}{c} c_k \\ d_k\end{array}\right)=\frac{1}{2}(1+i)\xi_n^{-\mu(x,y)}\left(\begin{array}{c} x'+y'\zeta_{2n}^k\\ x'-y'\zeta_{2n}^k\end{array}\right)
\]
and so
\[
\mu (c_k,d_k)-\mu\at{x,y}=\vp{1-i}-\min\at{\vp{x'+\zeta_{2n}^k y'},\vp{x' -\zeta_{2n}^k y'}}.
\]
To prove the Lemma we have to show that the right-hand side of this expression is strictly negative for some $k$.  To this end, it is sufficient to show that $\vp{x'+\zeta_{2n}^k y'} >\vp{1-i}=\vp{2}/2$. To see this, observe that
\[
\vp{\left(x'+\zeta_{2n}^k y'\right)+\left(x'-\zeta_{2n}^k y'}\right)=\vp{2x'}=\vp{2}
\]
and so if $\vp{x'+\zeta_{2n}^k y'} > \vp{2}/2$ then \eq{prop2} implies $\vp{x'-\zeta_{2n}^k y'} > \vp{2}/2$ as well.
	
Finally, to show that $\vp{x'+\zeta_{2n}^k y'} > \vp{2}/2 $ (for some $k$) it is sufficient to show
\begin{equation}
\vp{x' \md +\zeta_{2n}^k y' \md} > \vp{2}/2
\label{eq:suffcond}
\end{equation}
 (for some $k$). To see why this is sufficient, use \eq{padic} and the fact that
\[
 \at{x'+\zeta_{2n} y'} \md =\at{x' \md +\zeta_{2n}^k y' \md} \md.
\]
Now let $a=x' \md$ and $b= y' \md$. Since $\vp{x'}=\vp{y'}=0$, equation \eq{padic} shows that
\[
\vp{a}=\vp{b}=0.
\]
Furthermore
\[
\at{\abss{a} + \abss{b}} \md = \abss{x'} + \abss{y'}\md=0
\]
The last equality follows from the fact that $\abss{x'}+\abss{y'}\in 2\z$  which follows from $\vp{\abss{x'}+\abss{y'}}>\vp{2}$ (which is shown in \eq{absxy}). The pair $(a,b)$ therefore satisfies the hypotheses of \lemm{finite} and applying that Lemma we get that there exists an integer $k$ satisfying \eq{suffcond}.
 \qed

\begin{figure}
\begin{lstlisting}
check := function(n)
  C<w>:=CyclotomicField(2*n);
  CI:=Integers(C);
  PrimeIdealsAboveTwo := Decomposition(CI,2);
  assert(#PrimeIdealsAboveTwo eq 1 );
  p:=Place(PrimeIdealsAboveTwo[1,1]);
  twoValuation := func< x | Valuation(x,p) >;
  Z:=RingOfIntegers();
  mod2 := func< x | CI![ Z!y mod 2 : y in Eltseq(x) ]>;
  c := func< u | CI!Min([ Eltseq(mod2(u*w^k)) : k in [1..2*n]]) >;
  Elt := func< a | CI![x : x in a] >;
  sprime:=Set([ c(Elt(x)) : x in CartesianPower([0,1],Degree(CI)) | twoValuation(Elt(x)) eq 0 ]);
  abss := func< x | x*ComplexConjugate(x) >;
  S:=[ [x,y] : x in sprime, y in sprime | mod2(abss(x)+abss(y)) eq 0 ];
  checkcase := func< u,v | exists{ true : k in [1..2*n] | twoValuation(mod2(u+v*w^k)) gt twoValuation(2)/2 } >;
  return &and[ checkcase(p[1],p[2]) : p in S ];
end function;
check(2),check(4),check(6),check(8),check(12);
\end{lstlisting}
\caption{\label{fig:script} MAGMA script that verifies that \lemm{finite} is true. This script can be executed online at \url{http://magma.maths.usyd.edu.au/calc/}. }
\end{figure}


\section{Proof of \theo{phasethm}}\label{sec:phthm}

Recall that $n$ is even, $\ringR_n=\Z[\zeta_{2n},1/2]\subset\C$, and $T_n=\ringR_n\cap\R$. Write $U_R$ and $U_T$ for the unit groups of $\ringR_n$ and $T_n$, respectively.

\phases*
\vspace{.1in}

The central idea of the proof of this theorem is due to Shastri (\cite{Sh}, Theorem~1.1).  However, the details work out somewhat differently in our situation than in hers.

\vspace{.1in}

\noindent
{\it Proof:} \/ First, note that the set of roots of unity in $\ringR_n$ is the same as the set roots of unity in $\Q(\zeta_{2n})$, which is precisely the set of powers of $\zeta_{2n}$.  Thus, since the set $S$ is a group under multiplication, if we show that $S$ is finite, it must consist precisely of the roots of unity described in the statement of the theorem.

It remains to establish that any $r=a+bi\in \ringR_n$ satisfying $|r|=1$ is a root of unity if and only if there is an integer $t$ such that $2^t\equiv -1\pmod s$. For this part of the proof we may assume without loss of generality that $n$ is a multiple of $4$, since $\Z[\zeta_{2n},1/2]$ is contained in $\Z[\zeta_{8n},1/2]$, and the value of $s$ is the same for $2n$ and $8n$.  For the rest of the proof we therefore assume that $\ringR_n=\Z[\zeta_{2n},1/2]$, where $n$ is a multiple of $4$.

Let $T_n$ be the ring $T_n=\ringR_n\cap\R$ and let $K_n$ be the fraction field of $\ringR_n$. Since $a$ and $b$ are elements of $T_n$ satisfying $a^2+b^2=1$, $r=a+bi$ is an element of the kernel of the norm map $N\colon K_n\to K_n\cap\R$, given by \[N(a+bi)=(a+bi)(a-bi)=a^2+b^2=|a+bi|^2\]
Note that $N$ is a multiplicative homomorphism from $U_R$ to $U_T$.

The Dirichlet Unit Theorem computes the unit group of rings like $\ringR_n$ and $T_n$.  To state it, we will need some preliminaries about embeddings of $\ringR_n$ in $\C$.  For any number field $L$, let $r_1$ be the number of homomorphisms from $L$ to $\R$, and let $r_1+2r_2$ be the number of homomorphisms from $L$ to $\C$.  (The extra coefficient $2$ comes from the fact that complex embeddings come in conjugate pairs, if they are not real embeddings.)  It is known (see for example \cite{N}, page 30, the discussion before Proposition~5.1) that $r_1+2r_2=[L:\Q]$.

We reproduce here the specialization of the general Dirichlet Theorem to the case we need -- see \cite{N} for the general case.

\begin{thm}[\cite{N}, Proposition~VI.1.1]\label{thm:dirichlet}
Let $L$ be a number field.  Let $A$ be the ring of integers of $L$.  The group of units of the ring $A[1/2]$ is isomorphic to $\mu\times\Z^{r_1+r_2-1+y}$, where $\mu$ is the group of roots of unity in $L$, and $y$ is the number of prime ideals of $A$ containing $2$.
\end{thm}

Note that $\ringR_n$ is the ring $A[1/2]$ with $A=\Z[\zeta_{2n}]$, and $T_n$ is the ring $(A\cap\R)[1/2]$, so the theorem applies to both rings.

\begin{dfn}
Let $M$ be an abelian group isomorphic to $G\times\Z^m$, where $G$ is a finite group and $m$ is a non-negative integer.  The {\em rank} of $M$ is defined to be $m$.
\end{dfn}

Note that isomorphic groups have the same rank, and that every finitely generated abelian group is isomorphic to a group of the form $G\times\Z^m$.  Intuitively, an abelian group of rank $m$ is like a vector space of dimension $m$.  In particular, if $\phi\colon M_1\to M_2$ is a homomorphism of abelian groups of rank $m_1$ and $m_2$, respectively, then $\mbox{rank}(\ker\phi)+\mbox{rank}(\im\phi)=\mbox{rank}(M_1)=m_1$.

Since $\ringR_n$ cannot be embedded in $\R$ (it contains $i$), all of its complex embeddings are non-real.  The Dirichlet Unit Theorem says that the the group $U_R$ of units of $\ringR_n$ has rank $L+[\Q(\zeta_{2n}):\Q]/2$, where $L$ is the number of prime ideals of $\Z[\zeta_{2n}]$ that contain $2$.  The ring $T_n$, by contrast, admits no embeddings in $\C$ that are not contained in $\R$, so the group $U_T$ of units of $T_n$ has rank $\ell+[\Q(\zeta_{2n})\cap\R:\Q]=\ell+[Q(\zeta_{2n}):\Q]/2$, where $\ell$ is the number of prime ideals of $\Z[\zeta_{2n}]\cap\R$ that contain $2$:
\begin{align*}
\mbox{rank}(U_R) &= L+[\Q(\zeta_{2n}):\Q] \\
\mbox{rank}(U_T) &= \ell+[\Q(\zeta_{2n}):\Q]
\end{align*}

For every $b\in U_T$, we have $N(b)=|b|^2=b^2\in N(U_R)$.  By the Dirichlet Unit Theorem, if $m=\mbox{rank}(U_T)$, then there are units $b_1,\ldots,b_m$ such that the group $U_T$ is the set $\{wb_1^{a_1}\ldots b_m^{a_m}\}$ where $w$ is a root of unity and the $a_i$ are arbitrary integers.  The image $N(U_R)$ contains the set $\{w^2(b^2_1)^{a_1}\ldots (b^2_m)^{a_m}\}$, which also has rank $m$.  Since $N(U_R)$ is a subgroup of $U_T$, it therefore has rank exactly equal to $m=\mbox{rank}(U_T)$:
\[\mbox{rank}(N(U_R))=\mbox{rank}(U_T)=m\]

If $|r|=1$, then $r$ is an element of the kernel of $N$.  If we can show that the kernel of $N$ (restricted to $U_R$) is finite if and only if $-1$ is a power of $2$ modulo $s$, then we are done.

The rank of $U_R$ is $L+[\Q(\zeta_{2n}):\Q]/2$.  The rank of $N(U_R)$ is equal to the rank of $U_T$, which is $\ell+[Q(\zeta_{2n}):\Q]/2$.  If $L=\ell$, then the rank of $U_R$ and the rank of $N(U_R)$ will be equal, so the rank of the kernel of $N$ must be zero, implying that the kernel of $N$ is finite.  If $L\neq\ell$, then the kernel of $N$ will have positive rank, and there will be an infinite number of elements of $\ringR_n$ of absolute value $1$, and in particular infinitely many of them will not be roots of unity.  Thus, we wish to prove that $L=\ell$ if and only if $-1$ is a power of $2$ modulo $s$.

Consider the ring $B=\Z[\zeta_s]$.  A minimal polynomial for $\zeta_{2n}$ over $B$ is $x^{2^k}+\zeta_s$.  This is because the degree of $\zeta_{2n}$ over $\Q$ is $\varphi(2n)=\varphi(2^{k+1}s)=2^k\varphi(s)=2^k[\Q(\zeta_s):\Q]$, so $x^{2^k}+\zeta_s$ has the smallest possible degree amongst nonzero polynomials with coefficients in $\Z[\zeta_s]$ that have $\zeta_{2n}$ as a root.

If $P$ is any prime ideal of $\Z[\zeta_s]$ with $2\in P$, then modulo $P$, we have (for some $z'\in\Z[\zeta_s]/P$) $x^{2^k}+z\equiv(x+z')^{2^k}\pmod P$.  Prime ideals of $\Z[\zeta_{2n}]$ containing $P$ are in one-to-one correspondence with irreducible factors of $x^{2^k}+z$ modulo $P$.  Thus, for every prime ideal $P$ of $\Z[\zeta_s]$ with $2\in P$, there is exactly one prime ideal of $\Z[\zeta_{2n}]$ containing $P$.  Since any ideal of $\Z[\zeta_{2n}]$ containing $2$ must contain exactly one prime ideal of $\Z[\zeta_s]$ containing $2$, we see that the prime ideals of $\Z[\zeta_{2n}]$ containing $2$ are in one-to-one correspondence with the prime ideals of $\Z[\zeta_s]$ containing $2$.

Let $\gamma_{n}=\zeta_{n}+\zeta_{n}^{-1}$.  We will now show that the prime ideals of $\Z[\gamma_s]$ containing $2$ are in one-to-one correspondence with prime ideals of $\Z[\gamma_{2n}]$ containing $2$.

To do this, we will show that for any integer $c\geq 3$, the prime ideals of $\Z[\gamma_c]$ containing $2$ are in one-to-one correspondence with the prime ideals of $\Z[\gamma_{2c}]$ containing $2$.  A simple induction will yield the desired result.

If $c$ is odd, then $\Z[\gamma_c]=\Z[\gamma_{2c}]$ since $\gamma_c=-\gamma_{2c}$.  Thus, we assume that $c$ is even.

Notice that $\gamma^2_{2c}=(\zeta_{2c}+\zeta_{2c}^{-1})^2=\zeta_c+\zeta_c^{-1}+2$, so:
\[\gamma^2_{2c}=\gamma_c+2\]
A minimal polynomial for $\gamma_{2c}$ over $\Q(\gamma_c)$ is therefore $x^2-(\gamma_c+2)$.  If $P$ is a prime ideal of $\Z[\gamma_c]$ with $2\in P$, then $x^2-(\gamma_c+2)$ is a square modulo $P$ (indeed, it's congruent to $(x-\gamma_{2c})^2$.  Since the prime ideals of $\Z[\gamma_{2c}]$ containing $P$ are in one-to-one correspondence with the irreducible factors of $x^2-(\gamma_c+2)$ modulo $P$, there is exactly one prime ideal of $\Z[\gamma_c]$ containing $P$.  Thus, as in the case of $\zeta_{2n}$, we see that the prime ideals of $\Z[\gamma_{2c}]$ containing $2$ are in one-to-one correspondence with the prime ideals of $\Z[\gamma_c]$ containing $2$.

The rings $\Z[\gamma_s]$ and $\Z[\gamma_{2n}]$ have the same number of prime ideals containing $2$, and the rings $\Z[\zeta_s]$ and $\Z[\zeta_{2n}]$ have the same number of prime ideals containing $2$.  Thus, $L=\ell$ if and only if the rings $\Z[\gamma_s]$ and $\Z[\zeta_s]$ have the same number of prime ideals containing $2$.

Let $P$ be a prime ideal of $\Z[\gamma_s]$ with $2\in P$.  We will compute the number of prime ideals of $\Z[\zeta_s]$ containing $P$.

Such prime ideals correspond one-to-one with irreducible factors modulo $P$ of a minimal polynomial for $\zeta_s$ over $\Q(\gamma_s)$.  This polynomial is $x^2-\gamma_{s}x+1$.  Since $\gamma_s$ is a unit in the ring $\Z[\gamma_s]$, it cannot be an element of $P$.  Therefore, the polynomial $x^2-\gamma_sx+1$ cannot be a perfect square modulo $P$.  Thus, modulo $P$, the polynomial $x^2-\gamma_sx+1$ either has two coprime linear factors, or else it is irreducible.  In other words, there are two cases:
\begin{itemize}
\item There are two prime ideals $Q_1$, $Q_2$ of $\Z[\zeta_s]$ containing $P$, with $\Z[\zeta_s]/Q\cong\Z[\gamma_s]/P=F$ (``$P$ splits''), or else
\item There is a unique prime ideal $Q$ of $\Z[\zeta_s]$ containing $P$, satisfying $\Z[\zeta_s]/Q$ isomorphic to a degree two extension of $F$ (``$P$ is inert'')
\end{itemize}
We will show that $L=\ell$ if and only if $P$ is inert in $\Z[\zeta_s]$.

Let $Q$ be any prime ideal of $\Z[\zeta_s]$ containing $P$, and let $E$ be the field $\Z[\zeta_s]/Q$.  We see that $P$ splits if and only if $E=F$.

Let $m$ be the degree of $E$ over the field $\F_2$ with $2$ elements.  Then $m$ is the multiplicative order of $2$ modulo $s$.  To see this, notice that a field of order $2^m$ contains a primitive $s$th root of unity if and only if $2^m\equiv 1\pmod s$ -- elements of $\F_{2^m}$ are the roots of $x^{2^m}=x$, and $\zeta_s^s=1$.  The smallest positive integer $m$ satisfying that congruence is precisely the multiplicative order of $2$ modulo $s$.

The field $F$ is generated over $\F_2$ by the element $\gamma_s$ (or, more precisely, by its reduction modulo $P$).  We have $E=F$ if and only if there is no nontrivial element of $\mbox{Gal}(E/\F_2)$ that fixes $\gamma_s$.

The function $x\mapsto x^2$ is a generator of the Galois group of $E$ over $\F_2$, so every Galois conjugate (modulo $P$) of $\gamma_s$ is of the form $\zeta_s^{2^t}+\zeta_s^{-2^t}$ for some integer $t$.  If $a$ and $b$ are integers such that $\zeta_s^a+\zeta_s^{-a}=\zeta_s^b+\zeta_s^{-b}$ (modulo $P$), then $\zeta_s^a(1+\zeta_s^{b-a})=\zeta_s^{-b}(\zeta_s^{b-a}+1)$, implying that either $a\equiv -b\pmod s$, or else $\zeta_s^{b-a}\equiv 1\pmod P$, which is equivalent to $b\equiv a\pmod s$.  Thus, the $t$th Galois conjugate $\gamma_s^{2^t}$ of $\gamma_s$ equals $\gamma_s$ if and only if $2^t\equiv\pm 1\pmod s$.  Since $2^t\equiv 1\pmod s$ if and only if $t$ is a multiple of $m$, it follows that $\gamma_s$ is fixed by a nontrivial automorphism of $E$ if and only if $-1$ is a power of $2$ modulo $s$.

For every $P$, then, there is a unique prime ideal of $\Z[\zeta_s]$ containing $P$ if and only if $-1$ is a power of $2$ modulo $s$.  This conclusion is independent of $P$, so $L=\ell$ if and only if $-1$ is a power of $2$ modulo $s$.  Since there are elements $r\in S$ of infinite order if and only if $L\neq\ell$, we are done.  \qed

\section{Acknowledgments}
We thank Jean-Francois Biasse and Michele Mosca for helpful discussions. DG and DM were supported in part by NSERC. DG was supported in part by ARO. DG acknowledges funding provided by the Institute for Quantum Information and Matter, an NSF Physics Frontiers Center (NFS Grant PHY-1125565) with support of the Gordon and Betty Moore Foundation (GBMF-12500028). IQC is supported in part by the Government of Canada and the province of Ontario.
\bibliographystyle{plain}
\bibliography{qrotation}


\appendix
\section{Algebraic number theory glossary}\label{app:gloss}

In this paper, we use a number of terms and facts from algebraic number theory that may not be familiar to the reader.  We summarize them in this section.

\algint

By definition an algebraic integer is a root of some monic polynomial with integer coefficients; however, it is a fact that the \emph{minimal} polynomial of an algebraic integer over $\mathbb{Q}$ has integer coefficients.  The algebraic integers are a subring of $\mathbb{C}$; in particular, the sum and product of two algebraic integers is also an algebraic integer. Furthermore, if $z$ is a root of a monic polynomial which has algebraic integers as coefficients then $z$ is an algebraic integer.

\divisibility

\begin{dfn}
A number field is a finite algebraic extension of $\mathbb{Q}$, i.e., a field obtained by adjoining a finite set of algebraic numbers to $\mathbb{Q}$.
\end{dfn}

Every number field $K$ satisfies $K=\mathbb{Q}(\theta)$ for some algebraic number $\theta$; the degree of $K$ is the degree of the minimal polynomial of $\theta$.

\begin{dfn}
Let $K$ be a number field. The set $\mathcal{O}_K$ of all algebraic integers which lie in $K$ form a ring, called the ring of integers of $K$.
\end{dfn}
The ring of integers $\mathcal{O}_K$ of a number field is a Dedekind domain, which in particular implies that ideals admit unique prime factorization. That is, if $I\subset{\mathcal{O}_K}$ is a nonzero ideal and $I\neq \mathcal{O}_K$, there is a unique factorization (up to possible reordering of the factors):
\[
I=\prod_{j=1}^{k} Q_j
\]
where $Q_j$ are prime ideals.

\begin{dfn}
Let $K$ be a number field and $I,J\subset \mathcal{O}_K$ be two ideals which share no common factors in their prime factorizations. Then $I$ and $J$ are said to be coprime. Likewise, for any two elements $x,y\in \mathcal{O}_K$ we say $x$ and $y$ are coprime if $x\mathcal{O}_K$ and $y\mathcal{O}_K$ are coprime.
\end{dfn}
In fact, $I,J\subset\mathcal{O}_K$ are coprime if and only if $I+J=\mathcal{O}_K$.

\begin{dfn}
Let $K$ be a number field of degree $D$ and $I\subset \mathcal{O}_K$ be a nonzero ideal. The norm $N(I)$ of $I$ is equal to the number of elements of the quotient ring $\mathcal{O}_K/I$.
\end{dfn}
Note that the norm of a nonzero proper ideal (i.e.,  a nonzero ideal which is not equal to $\mathcal{O}_K$) is at least $2$.

The key property of the norm is that it is multiplicative, i.e., $N(IJ)=N(I)N(J)$. For a principal ideal generated by an element $a\in \mathcal{O}_K$ the norm can be equivalently expressed as follows.  Let $c$ be the constant coefficient of the monic minimal polynomial of $a$, and suppose this polynomial has degree $d$. Then $N(a\mathcal{O}_K)=|c|^{\frac{D}{d}}$ (here $d$ always divides $D$).  Since the constant coefficient of the monic minimal polynomial of $a$ is, up to sign, the product of the Galois conjugates of $a$, this means that we can also express the norm of a principal ideal as $N(a\mathcal{O}_K)=|a_1\ldots a_d|^{\frac{D}{d}}$.    (The Galois conjugates of an algebraic number $a$ are the algebraic numbers $\sigma(a)$, where $\sigma$ is a homomorphism from $\Q(a)$ to $\C$.  These are precisely the roots of the minimal polynomial of $a$.)

Note that the norm of an algebraic number will change depending on the field $K$.  In particular, if $L$ is a number field containing $K$, then $N(\alpha\mathcal{O}_L)=N(\alpha\mathcal{O}_K)^{[L:K]}$.  In particular, note that the norm of the ideal $2\mathcal{O}_K$ is equal to $2^{[K:\Q]}$.

If $a\mathcal{O}_K, b\mathcal{O}_K\subset \mathcal{O}_K$ have norms which are relatively prime, then $a$ and $b$ are coprime. To see this, write prime factorizations
\[
a\mathcal{O}_K=\prod_{j=1}^{k_1} Q_j \qquad b\mathcal{O}_K=\prod_{j=1}^{k_2} P_j
\]
where the norm of each of the prime factors is an integer $\geq 2$. Using multiplicativity we see that $P_j \neq Q_i$ for all $i$ and $j$ since otherwise $N(P_j)$ would divide both $N(a\mathcal{O}_K)$ and $N(b\mathcal{O}_K )$.

For further details about the norm of an ideal, see for example \cite{N,AW}.

\begin{dfn}
Let $\mathcal{O}_K$ be the ring of integers in a number field $K$ of degree $d$ over $\Q$.  An integral basis for $\mathcal{O}_K$ over $\Z$ is a set $\{r_1,\ldots,r_d\}$ such that every element of $\mathcal{O}_K$ can be uniquely expressed as an integer linear combination of $\{r_1,\ldots,r_d\}$.  In other words, for every $\alpha\in\mathcal{O}_K$, there exist unique integers $a_1,\ldots,a_d$ such that $\alpha=a_1r_1+\ldots +a_dr_d$.
\end{dfn}

\end{document}